\documentclass[12pt, prd, aps, showpacs, superscriptaddress, preprint, floatfix]{revtex4-1}
\usepackage[english]{babel}
\usepackage{float}
\usepackage{amsmath,amssymb,graphicx,textcomp}
\usepackage{mathtools}
\usepackage{mathrsfs}
\usepackage{hyperref}
\usepackage{subfigure}
\usepackage{slashed}

\newcommand\op{{\mathrm{op}}}
\newcommand\src{{\mathrm{src}}}
\newcommand\snk{{\mathrm{snk}}}

\newcommand\sep{{\mathrm{sep}}}

\def\simle{
    \mathrel{\rlap{\raise 0.511ex
        \hbox{$<$}}{\lower 0.511ex \hbox{$\sim$}}}}

\begin{document}

\title{Lattice Calculation of Hadronic Light-by-Light Contribution to the Muon
Anomalous Magnetic Moment}

\newcommand{\RBRC}{
  RIKEN BNL Research Center,
  Brookhaven National Laboratory,
  Upton, New York 11973,
  USA}

\newcommand{\UCONN}{
  Physics Department,
  University of Connecticut,
  Storrs, Connecticut 06269-3046,
  USA}

\newcommand{\NAGOYA}{
  Department of Physics,
  Nagoya University,
  Nagoya 464-8602,
  Japan}

\newcommand{\NISHINA}{
  Nishina Center,
  RIKEN,
  Wako, Saitama 351-0198,
  Japan}

\newcommand{\BNL}{
  Physics Department,
  Brookhaven National Laboratory,
  Upton, New York 11973,
  USA}

\newcommand{\CU}{
  Physics Department,
  Columbia University,
  New York, New York 10027,
  USA}

\newcommand{\KEK}{
  KEK Theory Center,
  Institute of Particle and Nuclear Studies,
  High Energy Accelerator Research Organization (KEK),
  Tsukuba 305-0801,
  Japan
}
\newcommand{\SOK}{
  School of High Energy Accelerator Science,
  The Graduate University for Advanced Studies (Sokendai),
  Tsukuba 305-0801,
  Japan
}

\author{Thomas Blum}
\affiliation{\UCONN}
\affiliation{\RBRC}

\author{Norman Christ}
\affiliation{\CU}

\author{Masashi Hayakawa}
\affiliation{\NAGOYA}
\affiliation{\NISHINA}

\author{Taku Izubuchi}
\affiliation{\BNL}
\affiliation{\RBRC}

\author{Luchang Jin}
\affiliation{\CU}

\author{Christoph Lehner}
\affiliation{\BNL}

\date{October 21, 2015}

\begin{abstract}
The quark-connected part of the hadronic light-by-light scattering contribution to the muon's anomalous magnetic moment is computed using lattice QCD with chiral fermions. We report several significant algorithmic improvements and demonstrate their effectiveness through specific calculations which show a reduction in statistical errors by more than an order of magnitude.  The most realistic of these calculations is performed with a near-physical, $171$ MeV pion mass on a  $(4.6\;\mathrm{fm})^3$ spatial volume using the $32^3\times 64$ Iwasaki+DSDR gauge ensemble of the RBC/UKQCD Collaboration.
\end{abstract}

\maketitle

\section{Introduction}
\label{sec:intro}

New particles and interactions which occur at a very large energy scale $\Lambda$, above the reach of present-day accelerators, may be first discovered through their indirect effects at low energy.   A particularly promising low energy quantity that may reveal such effects is the anomalous moment of the muon.  This ``anomalous'' difference $g_\mu-2$ between the muon's gyromagnetic ratio $g_\mu$ and the Dirac value of 2 for a non-interacting particle can receive contributions from such new high energy  phenomena, contributions which are suppressed by the ratio of the squares of the energy scales  $(m_\mu/\Lambda)^2$ and the strength of the coupling of these new phenomena to the muon.  (Here $m_\mu=105$ MeV is the mass of the muon.)  The known couplings of the muon are its relatively weak interaction with the photon, the $W^\pm$, $Z$ and Higgs bosons, which can be accurately described by perturbation theory.  This implies even very small differences between $g_\mu-2$ and the predictions of the standard model can be recognized, making $g_\mu-2$ an attractive place to search for new, beyond-the-standard-model phenomena~\cite{Miller:2007kk}.

In fact, the use of $g_\mu-2$ to search for new phenomena has reached a very high level of precision.   This quantity has been measured with an accuracy of 0.54~ppm~\cite{Bennett:2006fi} and the corresponding theoretical calculations have achieved a similar level of precision.  The present status of experiment and theory is summarized in Tab.~\ref{tab:g-2-th-ex}.  As this table shows there is at present a 3 standard deviation discrepancy between the experimental result and the standard model prediction.   This discrepancy provides strong motivation both for new experiments, which are either underway or planned at Fermilab (E989) and J-PARC (E34) with a targeted precision as small as 0.14 ppm, and for a reduction in the theoretical errors.

\begin{figure}[!htbp]
  \resizebox{0.35\columnwidth}{!}{\includegraphics{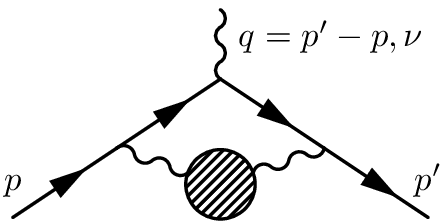}}
  \hskip 0.1\columnwidth
  \resizebox{0.35\columnwidth}{!}{\includegraphics{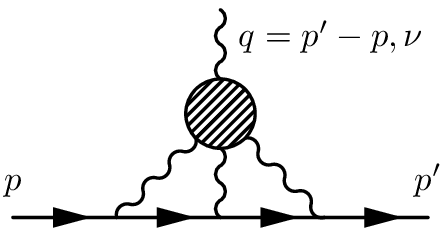}}
  \caption{Feynman diagrams depicting the hadronic vacuum polarization (left) and hadronic light-by-light scattering (right) contributions to $g_\mu-2$.}
\label{fig:hvp-hlbl}
\end{figure}

The two components of the theoretical calculation with the largest errors involve couplings to the up, down and strange quarks: the hadronic vacuum polarization (HVP) and hadronic light-by-light scattering (HLbL).   These are the first cases in which the effects of the strong interaction enter the determination of $g_\mu-2$.  The HVP effects enter beginning at order $\alpha^2$ while those from HLbL are of order $\alpha^3$, where $\alpha = 1/137.036$ is the fine structure constant.  These two types of contributions are shown in Fig.~\ref{fig:hvp-hlbl} and, because of the strong interactions of the quarks, these quantities must be evaluated using methods which treat the strong interactions non-perturbatively.

\begin{table}[h]
  \begin{center}
    \begin{tabular}{lrr}
    \hline\hline
      SM Contribution & Value $\pm$ Error & Ref\\
      \hline
      QED (incl. 5-loops) & $116584718.951 \pm 0.080$ & {\cite{Aoyama:2012wk}}\\
      HVP LO &  $6949 \pm 43$ & {\cite{Hagiwara:2011af}}\\
      HVP NLO & $- 98.4 \pm 0.7$ & {\cite{Hagiwara:2011af,Kurz:2014wya}}\\
      HVP NNLO & $12.4 \pm 0.1$ & {\cite{Kurz:2014wya}}\\
      HLbL & $105 \pm 26$ & {\cite{Prades:2009tw}}\\
      Weak (incl. 2-loops) & $153.6 \pm 1.0$ & {\cite{Gnendiger:2013pva}}\\
      \hline
      SM Total   (0.51 ppm)& $116591840 \pm 59$ & {\cite{Aoyama:2012wk}}\\
      \hline
      Experiment (0.54 ppm) & $116592089 \pm 63$ & {\cite{Bennett:2006fi}}\\
      Difference
      ($\ensuremath{\operatorname{Exp}}-\ensuremath{\operatorname{SM}}$) &
      $249 \pm 87$ & {\cite{Aoyama:2012wk}}\\
      \hline\hline
    \end{tabular}
  \end{center}
  \caption{Comparison between experiment and the standard model prediction for $(g_\mu-2)/2$ (in units of $10^{- 11}$). Other recent analyses~\cite{Davier:2010nc, Hagiwara:2011af} give similar values for the difference between experiment and standard model theory.  Note that the HVP NNLO contribution is not included in the standard model totals, while LO, NLO and NNLO indicates leading order, next-leading order and next-next-leading order.}
\label{tab:g-2-th-ex}
\end{table}

The strong interaction contribution to HVP can be determined directly from the experimentally measured cross-section for the single photon $e^+$--$\,e^-$ annihilation into hadrons using a dispersion relation --- a well-developed method with fractional percent errors.   These same non-perturbative strong interaction effects can be determined using lattice QCD~\cite{Blum:2002ii} but accuracy comparable to that obtained from  experimentallly measured $e^+$--$\,e^-$ annihilation has yet to be achieved.  The determination of the HVP contribution by both methods is an active area of research~\cite{Blum:2013xva, Benayoun:2014tra} and further reduction of these errors is expected.

The HLbL contribution is less well studied and is the topic of this paper.  Unlike the HVP case, it is presently not known how to determine the HLbL contribution from experimental data and dispersion relations, although progress is being made in this direction~\cite{Colangelo:2014dfa, Colangelo:2014pva, Pauk:2014jza, Pauk:2014rfa, Colangelo:2015ama}.  The HLbL contribution to $g_\mu-2$ has been evaluated in model calculations~\cite{Prades:2009tw,Benayoun:2014tra} whose errors can not be systematically improved and whose estimates, which are used in Tab.~\ref{tab:g-2-th-ex},  are approximately of the same size as the discrepancy between the standard model theory and experiment.

However, as demonstrated by Blum, Chowdhury, Hayakawa and Izubuchi~\cite{Blum:2014oka}, this quantity can be calculated from first principles using the methods of lattice QCD.  Unfortunately, as their calculation also demonstrates, even the most accessible quark-connected part of the HLbL contribution is a challenging task for lattice QCD especially if physical quark masses and realistically large volumes are to be used.  The more difficult disconnected parts, while also accessible to a first-principles lattice calculation, will be even more demanding.

In the present paper we develop a series of significant improvements to the methods used in the paper of Blum {\it et al.} and demonstrate their effectiveness with several calculations, including one at a much smaller, 171 MeV pion mass in a large $(4.6\;\mathrm{fm})^3$ spatial volume.  These improvements are described as a series of steps which reduce both systematic and statistical errors while giving greater insight into the quantity being computed.   Throughout this paper we will focus on the connected HLbL amplitude which will be abbreviated as cHLbL.

In the first step (Sec.~\ref{sec:stochastic_field}) we move from the non-perturbative treatment of QED used in Ref.~\cite{Blum:2014oka} to one in which explicit stochastic electromagnetic fields are introduced which generate only the three photon propagators which appear in the $O(\alpha^{3})$ HLbL amplitude.   This avoids entirely $O(\alpha^{2})$ statistical errors as well as the unwanted $O(\alpha^{4})$ contributions present in the earlier, non-perturbative approach to QED.

In the second step (Sec.~\ref{sec:exact_props}) these stochastically generated photon propagators are replaced by the analytic propagators which they approximate.   Of course, when making such a replacement we lose the important benefit offered by the stochastic approach:  when a photon propagator is generated as the average of a product of stochastic fields, the complete amplitude can be written as the product of separate factors, one containing the source field and the other the sink field.  It is only when this product is averaged over the stochastic field that a coupling between these factors is introduced.  A calculation of (volume$)^2$ difficulty is replaced by the average of products, each of only (volume$)^1$ difficulty.

\begin{figure}[!htbp]
  \begin{center}
    \resizebox{0.35\columnwidth}{!}{\includegraphics{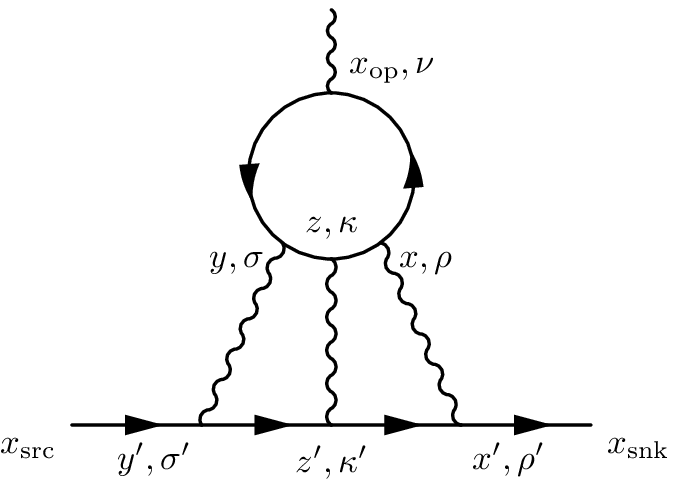}}
    \hskip 0.1\columnwidth
    \resizebox{0.35\columnwidth}{!}{\includegraphics{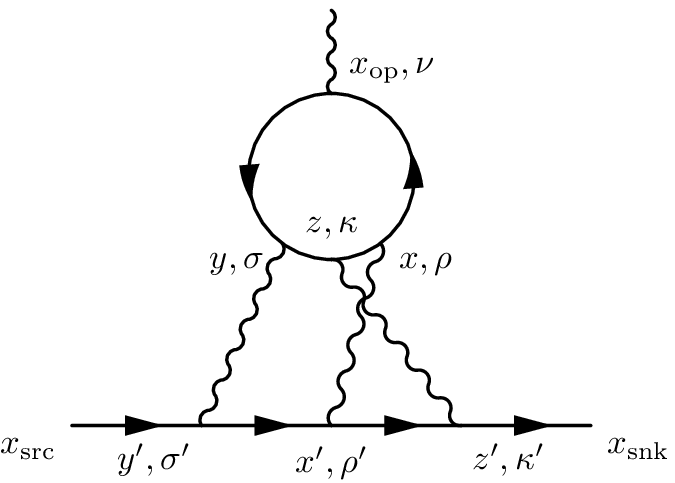}}
  \end{center}
  \caption{Hadronic light-by-light diagrams. There are four additional diagrams resulting from further permutations of the photon vertices on the muon line.}
\label{fig:hlbl}
\end{figure}

We overcome the (volume$)^3$ problem that results when three analytic photon propagators are introduced, by stochastically summing over the locations where two of the photons couple to the internal quark line.  For example, referring to Fig ~\ref{fig:hlbl} we might evaluate each amplitude for a series of random space-time locations of the vertices at $x$ and $y$ and then stochastically sum over $x$ and $y$.  This replacement of a stochastic evaluation of the $4L^3 T$-dimensional integral over the electromagnetic field by the much simpler stochastic evaluation of the 8-dimensional sum over two electromagnetic vertices dramatically simplifies the calculation.  Here $L$ and $T$ are the spatial and temporal extents of the lattice volume.  Since the two vertices appear on the same closed quark loop, the amplitude being evaluated will fall exponentially as $x$ and $y$ are separated beyond $\approx 1$ fm, a fact that can be exploited when choosing the distribution according to which $x$ and $y$ are generated.

As is shown in Appendix~\ref{sec:conserved-vs-local}, the short distance properties of these HLbL graphs require that at least one of the currents which couple to the internal quark line must be a conserved lattice current if the resulting amplitude is to have a simple continuum limit with no need to subtract a contact term.  The conservation of the external current implies that this amplitude vanishes in the limit that $q\to0$, the limit needed to evaluate $g_\mu-2$.  The third algorithmic improvement (Sec.~\ref{sec:WI_per_config}) that we explore is making a choice of graphs so that this vanishing behavior in the $q\to0$ limit occurs for each QCD gauge configuration.  If this approach is adopted then both the signal and the noise will vanish in this limit.

The fourth algorithmic development (Sec.~\ref{sec:moment}) resolves the difficulty of evaluating the limit $q^2\to0$ for an amplitude which is proportional to $q$ in finite volume.  In such a case the amplitude would normally be evaluated at the smallest, non-zero lattice momentum $2\pi/L$ and the limit $q^2\to0$ achieved only in the limit of infinite volume (or by extrapolation from non-zero $q^2$).  Here we introduce a position-space origin related to the choice of $x$ and $y$ and show that a simple, spatial first moment of the finite-volume, current matrix element between zero-momentum initial and final muons will yield the $q^2=0$ anomalous magnetic moment:
\begin{equation}
 (g_\mu-2)_{\mathrm{cHLbL}}\frac{-e}{4m_\mu} \vec\sigma_{s's} =  \frac{1}{2}\int d^3 r \left\{\vec r \times
                     \bigl\langle\mu(s')\bigl|\,\vec J(\vec r)\, \bigr|\mu(s)\bigr\rangle_{\mathrm{cHLbL}}\right\}.
\end{equation}
Here $\vec\sigma$ is a vector formed from the three Pauli matrices, $s$ and $s'$ are the initial and final spin indices and the label cHLbL indicates that only the quark-connected, HLbL amplitude is being considered.  The relation between the initial and final states, the electromagnetic current $\vec J(\vec r)$ and the volume will be carefully specified below.

The paper is organized as follows.  In Sec.~\ref{sec:strategy} we describe in greater detail the algorithmic improvements outlined above. Section~\ref{sec:studies} contains the numerical results that demonstrate these new methods.  Two results are of particular interest. The first is a value for the quark-connected HLbL contribution:
\begin{equation}
  \frac{(g_\mu-2)_{\mathrm{cHLbL}}}{2} = (0.1054 \pm 0.0054) (\alpha/\pi)^3 = (132.1 \pm 6.8)\times 10^{-11},
\label{eq:result}
\end{equation}
obtained with a 171 MeV pion mass and $(4.6\; \mathrm{fm})^3$ volume, the most realistic lattice QCD calculation of this quantity to date.  While it is premature to compare this result with experiment or model calculations because the errors arising from finite-volume, finite-lattice spacing and the unphysical quark and muon masses are not yet controlled, the 5\% statistical error suggests that this calculation is now within the reach of the methods of lattice QCD.  The second result of special interest is for pure QED where a muon loop instead of a quark loop appears.  In this case all of the diagrams are connected so our calculation should give the complete result.  Here we work at $q^2=0$ and examine three values for the lattice spacing $a$ (actually three values of $m_\mu a$) and three physical volumes.  We use the three choices of lattice spacing to extrapolate to the continuum limit and are then able to recognize a $1/L^2$ dependence on the spatial extent $L$ of the volume.  Using this form to extrapolating to $L\to\infty$, we obtain a continuum and infinite volume limit which is consistent with the known, perturbative QED result.  A summary and outlook are given in Sec.~\ref{sec:conclusion}.  We should emphasize that as in Ref.~\cite{Blum:2014oka}, only the quark-connected HLbL contribution has been considered and the quark-disconnected diagrams, where two, three, or four quark loops couple to the external current and the three internal photon propagators, are not discussed.

\section{Evaluation Strategy}
\label{sec:strategy}

The anomalous magnetic moment of the muon is determined by the electromagnetic form factor $F_2(q^2)$  evaluated at $q^2=0$: $F_2 (0) = (g_\mu-2)/2 \equiv a_\mu$ where $a_\mu$ is known as the muon anomalous magnetic moment  and the usual form factors $F_1$ and $F_2$ appear in the decomposition of the matrix element of the electromagnetic current between the incoming and outgoing muon states:
\begin{eqnarray}
\langle \mu(\vec p\,')|J_\nu(0)|\mu(\vec p)\rangle & = & -e\overline{u}(\vec p\,') \Bigl(F_1 (q^2) \gamma_\nu
    + i \frac{F_2(q^2)}{4 m}[\gamma_\nu, \gamma_\rho] q_\rho \Bigr) u(\vec p),
\label{eq:vertex-func_1}
\end{eqnarray}
where $J_\nu(0)$ is the electromagnetic current, $|\mu(\vec p)\rangle$ and $|\mu(\vec p\,')\rangle$ the initial and final muon states, $u(\vec p)$ and $\overline{u}(\vec p\,') = u^\dagger(\vec p\,') \gamma^0$ are standard, positive-energy solutions of the Dirac equation and $-e$ the electric charge of the muon.   The states $|\mu(\vec p)\rangle$ and $|\mu(\vec p\,')\rangle$ are normalized as simple plane waves.  Thus, in finite volume their inner product will be given by $V\delta_{\vec p,\vec p\,'}$ while in infinite volume $(2\pi)^3 \delta( \vec p-\vec p\,')$ will result.

The matrix element in Eq.~\eqref{eq:vertex-func_1} can be obtained from a Euclidean-space lattice QCD calculation be evaluating a Euclidean-space Green's function containing a muon source and sink with definite incoming and outgoing momentum (here chosen to be $-\vec q/2$ and $\vec q/2$, respectively) in the limit of large time separation:
\begin{eqnarray}
\mathcal{M}_\nu(\vec q)  &=& \lim_{t_\src\to -\infty \atop t_\snk\to\infty}
      e^{E_{q/2}(t_\snk-t_\src)}  \sum_{\vec{x}_\snk, \vec{x}_\src}
      e^{-i\frac{\vec q}{2} \cdot (\vec{x}_\snk + \vec{x}_\src)}
      e^{i\vec{q}\cdot\vec{x}_\op} \mathcal{M}_\nu(x_\snk, x_\op, x_\src ),
\label{eq:M_p-src}
\end{eqnarray}
where $E_{q/ 2} = \sqrt{(q/ 2)^2 +m_\mu^2}$ and the amplitude $\mathcal{M}_\nu ( x_\src, x_\op, x_\snk)$ is given by the Euclidean-space Green's function
\begin{eqnarray}
-e \mathcal{M}_\nu ( x_\src, x_\op, x_\snk) &=& \bigl\langle\mu(x_\snk) J_\nu(x_\op)\overline{\mu}(x_\src)\bigr\rangle.
\end{eqnarray}
Here the operator $\overline{\mu}(x_\src)$ creates a muon at the space-time position $x_\src$,
$\mu(x_\snk)$ destroys a muon at the position $x_\snk$ and $J_\nu(x_\op)$ is the operator for the electromagnetic current.  For the general case discussed in this and the following paragraph, the fields $\overline{\mu}(x_\src)$ and $\mu(x_\snk)$ must be renormalized, a refinement which is not needed for the class of graphs which enter the HLbL contribution to $g_\mu-2$.  Note, the factor $e^{i\vec{q}\cdot\vec{x}_\op}$ has been introduced into Eq.~\eqref{eq:M_p-src} so that translational symmetry implies that $\mathcal{M}_\nu(\vec q)$ does not depend on the position $x_\op$.

Recognizing that the two Euclidean-time limits, $t_\src\to -\infty$ and $t_\snk\to\infty$ in Eq.~\eqref{eq:M_p-src} will project onto physical muon states, we can relate the form factors in Eq.~\eqref{eq:vertex-func_1} and the amplitude $\mathcal{M}_\nu(\vec q)$:
\begin{equation}
\left[\Bigl(\frac{-i\slashed q^+ +m_\mu}{2E_{q/2}}\Bigr)
   \Bigl(F_1 (q^2) \gamma_\nu + i \frac{F_2(q^2)}{4 m}[\gamma_\nu, \gamma_\rho] q_\rho \Bigr)
        \Bigl(\frac{-i\slashed q^- +m_\mu}{2E_{q/2}}\Bigr)\right]_{\alpha\beta}
 = \Bigl(\mathcal{M}_\nu(\vec q)\Bigr)_{\alpha\beta},
\label{eq:vertex-func_2}
\end{equation}
where for clarity we have explicitly introduced the spinor indices $\alpha$ and $\beta$ and the four- momenta have the form $q^\pm = (i E_{q/2}, \pm \vec{q}/2)$.

We now specialize to the cHLbL case of interest and its particular set of six graphs, two of which appear in Fig.~\ref{fig:hlbl}.  In this case, it will be convenient to express $\mathcal{M}_\nu ( x_\src, x_\op, x_\snk)$ as an explicit sum of an amplitude $\mathcal{F}_\nu(x, y, z, x_\op, x_\snk, x_\src)$ in which the locations of the other three photon-quark vertices, $x$, $y$ and $z$, indicated in Fig.~\ref{fig:hlbl}, appear:
\begin{equation}
\mathcal{M}_\nu ( x_\src, x_\op, x_\snk) = \sum_{x,y,z} \mathcal{F}_\nu(x, y, z, x_\op, x_\snk, x_\src).
\label{eq:hlbl-amp_1}
\end{equation}
The amplitude $\mathcal{F}_\nu(x, y, z, x_\op, x_\snk, x_\src)$ can then be written in terms of quark, muon and photon propagators:
\begin{eqnarray}
\mathcal{F}_\nu ( x, y, z, x_\op, x_\snk, x_\src) &=&
\nonumber \\
&& \hskip -1.0 in  - (- i e)^2 \sum_{q = u, d, s} (ie_q)^4
\Bigl\langle \mathrm{tr} \bigl[
  \gamma_{{\nu}} S_q \left( x_\op, x \right) \gamma_{\rho} S_q (x,
  z) \gamma_{\kappa} S_q (z, y) \gamma_{\sigma} S_q \left( y, x_\op
  \right) \bigr] \Bigr\rangle_{\text{QCD}} \nonumber\\
  && \hskip -0.5 in \cdot \sum_{x', y', z'} G_{\rho \rho'} (x, x') G_{\sigma \sigma'} (y,
  y') G_{\kappa \kappa'} (z, z') \nonumber \\
&&\hskip -0.0 in \cdot\Bigl[ S_\mu \left( x_{\text{snk}},
  x' \right) \gamma_{\rho'} S_\mu (x', z')\gamma_{\kappa'} S_\mu (z', y')
\gamma_{\sigma'} S_\mu \left( y', x_{\text{src}} \right) \nonumber \\
&&\hskip 0.25 in  +S_\mu \left( x_{\text{snk}}, z' \right)
  \gamma_{\kappa'} S_\mu (z', x') \gamma_{\rho'}S_\mu (x', y') \gamma_{\sigma'}
  S_\mu \left( y', x_{\text{src}} \right)
  \nonumber\\
  && \hskip 0.25 in + \text{four other permutations} \Bigr],
\label{eq:hlbl-amp_2}
\end{eqnarray}
where only the two sets of contractions shown in Fig.~\ref{fig:hlbl} are written explicitly.  For simplicity, Eq.~\eqref{eq:hlbl-amp_2} is written using local operators for each of the seven electromagnetic currents.  The electric charge of the muon is $-e$, while $e_u=2e/3$, $e_d=e_s=-e/3$ are the charges of the up, down and strange quarks.  The  brackets $\langle\ldots\rangle_{\mathrm{QCD}}$ indicate an average over the QCD gauge configurations which provide the background fields in which the quark propagators $S_q(x,y)$ are computed.  The quantities $G_{\sigma,\sigma'}(x,y)$ and $S_\mu(x,y)$ are photon and muon propagators respectively.  The polarization indices are shown explicitly on the photon propagators but $S_\mu$ and $S_q$ are $4\times4$ spinor matrices with the spin indices suppressed.  We use Euclidean-space conventions with the $\gamma$ matrices obeying $\{\gamma_\nu,\gamma_\rho\} = 2\delta_{\nu,\rho}$ as specified in Appendix~\ref{sec:conventions}

The six sums over the space-time volume which appear in Eqs.~\eqref{eq:hlbl-amp_1} and \eqref{eq:hlbl-amp_2} make this expression too computationally expensive to be evaluated directly and stochastic methods must be introduced if this quantity is to be computed with current computing resources.

\subsection{Stochastic electromagnetic field}
\label{sec:stochastic_field}

One standard stochastic method of including electromagnetic effects is to compute the charged fermion propagators in the background of stochastically generated QED gauge field configurations. If these gauge configurations are generated according to a discrete version of the Maxwell action, then averaging over these QED configurations will reproduce all photon exchange diagrams in exact analogy with the usual technique for including the gluon degrees of freedom in lattice QCD.   However, this method will include QED contributions to all orders in $\alpha$, beginning at order $\alpha^1$. Since we are only interested in $\mathcal{O} (\alpha^3)$ contributions corresponding to the diagrams in Fig.~\ref{fig:hlbl}, we must perform a carefully crafted subtraction to remove the lower order contributions while keeping $\alpha$ small to control the higher order contribution~\cite{Hayakawa:2005eq}.   This method has been successfully applied to obtain the first lattice QCD results for this cHLbL contribution and requires the evaluation of relatively few quantities because of the indirect treatment of most of the electromagnetic degrees of freedom.  However, as $\alpha$ is decreased to reduce the size of the unwanted $\alpha^4$ and higher order diagrams, we must deal with the lower order $\alpha^2$ terms which, although vanishing on average because of the subtractions which are performed, can still contribute to the stochastic noise.

\begin{figure}[!htbp]
  \begin{center}
    \resizebox{0.35\columnwidth}{!}{\includegraphics{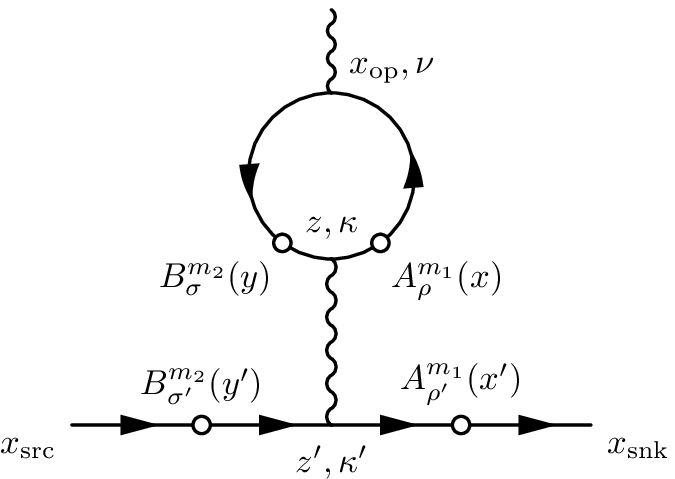}}
    \hskip 0.075\columnwidth
    \resizebox{0.35\columnwidth}{!}{\includegraphics{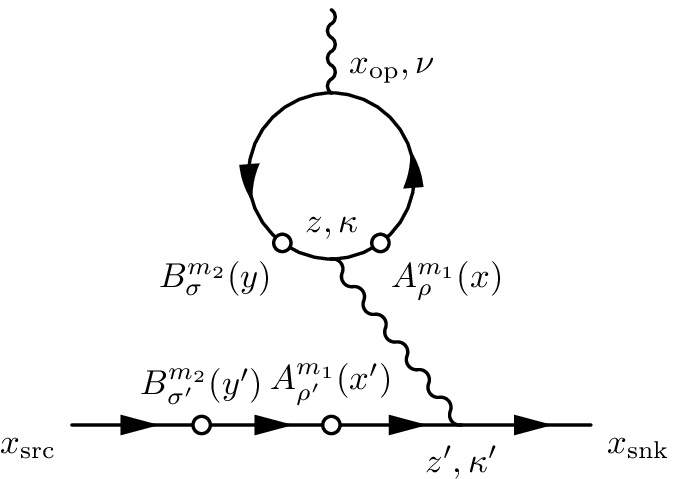}}
  \caption{Two of the six HLbL diagrams that result if a stochastic method is adopted to evaluate two of the three photon propagators which appear in Fig~\ref{fig:hlbl}.  The wavy line joining the muon line and the quark loop represents the exact photon propagator while the pairs of factors, $A^{m_1}_\rho(x)$,  $A^{m_1}_{\rho'}(x')$ and $B^{m_2}_\sigma(y)$,  $B^{m_2}_{\sigma'}(y')$ are the $m_1$ and $m_2$ elements of two independent ensembles of stochastic electromagnetic fields.}
\label{fig:hlbl_AB}
\end{center}
\end{figure}

In fact, stochastic methods can be used to directly evaluate the specific graphs of interest if one begins with an expression very similar to the $\alpha^3$ amplitude of interest given in Eq.~\eqref{eq:hlbl-amp_2}.  We can simply replace the photon propagators $G_{\rho,\rho'}(x,y)$ which appear in that equation by the product of two stochastic variables distributed so that the average of their product reproduces the target propagator:
\begin{equation}
\bigl\langle A_\rho(x) A_{\rho'}(y)\bigr\rangle_A = G_{\rho,\rho'}(x,y),
\label{eq:stochastic_photon}
\end{equation}
where $\bigl\langle\ldots\bigr\rangle_A$ represents an average over this ensemble of electromagnetic gauge fields.   An appealing implementation of this approach follows the original construction of Blum {\it et al.} and replaces only the photon propagators which couple to the left $(x)$ and right $(y)$ points along the quark line in Fig.~\ref{fig:hlbl} by stochastic fields while keeping an exact photon propagator which joins the quark line at the center point $z$.  The unwanted propagator joining the points $x$ and $y$ can be avoided if independent stochastic fields are used for the points $x$ and $y$.  If these two stochastic fields are written as $A_\rho(x)$ and $B_\sigma(y)$ then the two diagrams shown in Fig.~\ref{fig:hlbl} are simplified to those shown in Fig.~\ref{fig:hlbl_AB}.

With the introduction of these two stochastic field variables the evaluation of the amplitudes corresponding to the diagrams shown in Fig.~\ref{fig:hlbl_AB} is straight-forward.  Each product of two quark propagators joined by a stochastic field can be evaluated using the sequential source method.  For example, consider the quark propagator on the left side of the loop, coupling to the $B^{m_2}$ field in Fig.~\ref{fig:hlbl_AB}.  The location of the external current $x_\op$ can be used as a source allowing us to solve for the first propagator, $S_q(y,x_\op)$ which is found as a function of the sink position $y$.  This function can then be multiplied by $B_\sigma^{m_2}(y)\gamma_\sigma$ and the resulting function of $y$ used as a source for the second propagator which connects to the vertex $z$.  This same approach can be used to obtain the product of quark propagators joining $x_\op$ and $z$ as well as the two products of pairs of muon propagators needed to construct the muon line.  Finally the resulting two explicit functions of $z$ and $z'$ can be multiplied by the exact photon propagator connecting $z$ and $z'$ and the final sum over $z$ and $z'$ performed with $O(V\ln(V))$ operations by using the fast Fourier transform (FFT).

We should point out that while the discussion above is simplest if we use a fixed location, $x_\op$ for the external current vertex, in a practical calculation a sum over this position can be achieved by using a random source for the two propagators joined to $x_\op$, which is distributed over a possibly large space-time subvolume and will lead to a much improved signal-to-noise ratio.  In this standard method, arranging the noise source as a vector of independent random numbers for each site guarantees that after a noise average only the desired terms where the two propagators are joined to the same point will be non-zero.

An interesting enhancement that can be exploited when using this method is to compute the $2M$ sequential source propagators for the right- and left-hand quark propagators shown in Fig.~\ref{fig:hlbl_AB} separately, where the right-hand sequential source propagator incorporates stochastic field $A^{m_1}$ while the left-hand propagator contains $B^{m_2}$ for  $1 \le m_1,m_2 \le M$.  We can then compute the amplitude of interest for all $M^2$ pairs, effectively enhancing the statistical sample by a factor of $M$ with only the added cost of $M^2$ evaluations of the less expensive muonic part of the amplitude.  We refer to this approach as the $M^2$ method and present numerical results in Sec.~\ref{sec:M-squared}. These results suggest that the full statistical gain of a factor $M^2$ is realized.

Introducing specific, stochastic QED fields and using sequential source propagators solves the problem of lower order noise that will degrade a dynamical QED calculation in which amplitudes of lower order in $\alpha$ are removed by subtraction.  However, there is another very significant problem, which might be called the ``disconnected-diagram'' problem. If we were to replace all three photon propagators with pairs of stochastic QED fields obeying the condition given in Eq.~\eqref{eq:stochastic_photon}, the resulting diagram would usually be referred to as a disconnected diagram because the quark loop and the muon line are not joined by explicit propagators which decrease as their endpoints are separated.  For example, if we work with fixed spatial locations for $x_\src$, $x_\op$ and $x_\snk$ but allow the time separation $t_\sep$ between $x_\src$ and $x_\snk$ to grow (to project onto the muon ground state), each stochastic field will contribute unsuppressed noise from any point along the muon line.  For the case when three stochastic photon propagators are used, these stochastic fluctuations will cause the statistical error to grow as $t_\sep^{3/2}$, where we estimate the stochastic noise by averaging the square of the product of the three fields evaluated on the muon line, giving a result for the square of the noise which grows proportional to $t_\sep^3$.  This noise problem will become even more severe if we work in a large spatial volume and use a wall source for the initial and final muon and a random wall source, also at fixed time, for the external current in an attempt to exploit a finite volume average.  The result will be a statistical error which should grow as $L^3$ assuming that $L$ and $T$ are of approximately the same size.  (This estimate comes from combining the factor of $T^3$ obtained in the estimate above with a factor of $L^3$ resulting from the integration of $x_\op$ over the $L^3$ volume contributing at a fixed time, implying an error whose square will grow as $T^3L^3$.)   If one exact photon propagator is introduced as discussed above, these effects are reduced, but the resulting statistical error will still grow as $L^2$ since the removal of one of the stochastic fields evaluated on the muon line will reduce the $T$-dependence of square of the error from $T^3$ to $T^2$ and the presence of the explicit photon propagator joining the quark loop and muon line will reduce the contribution from the integration over $x_\op$ from $L^3$ to $L^2$.

\subsection{Exact photon propagators}
\label{sec:exact_props}

To completely avoid this disconnected-diagram problem, we need to use an explicit, free-field formula for each of the three photon propagators and introduce the necessary stochastic sampling in a different way.   Fortunately, this is not difficult and will result in statistical noise that will remain finite, even in the infinite volume limit.  This new approach to the HLbL calculation is the topic of this section.  As suggested above, it is not possible to evaluate Eq.~\eqref{eq:hlbl-amp_1} and \eqref{eq:hlbl-amp_2} without approximations even on a single QCD configuration, so we introduce randomness in a different way which, as we will see, leads to statistical fluctuations which are much more easily controlled.

This approach can be best presented if we express the cHLbL amplitude $\mathcal{M}_\nu(\vec q)$ as an explicit sum over the three additional space-time vertices $x$, $y$ and $z$ at which the internal photon lines couple to the quark line, in analogy with Eqs.~\eqref{eq:hlbl-amp_1} and \eqref{eq:hlbl-amp_2}:
\begin{eqnarray}
\mathcal{M}_\nu(\vec q) &=& e^{i\vec{q}\cdot\vec{x}_\op} \sum_{x,y,z} \mathcal{F}_\nu(\vec q, x, y, z, x_\op),
\label{eq:M_p-src-space}
\end{eqnarray}
where the factor of $e^{i\vec{q}\cdot\vec{x}_\op}$ has been introduced so that $\mathcal{M}_\nu(\vec q)$ will not depend on $x_\op$ and the amplitude $\mathcal{F}_\nu(\vec q, x, y, z, x_\op)$ is related to the similar point-source/point-sink quantity defined in Eq.~\eqref{eq:hlbl-amp_2} by:
\begin{eqnarray}
\mathcal{F}_\nu(\vec q, x, y, z, x_\op) &=& \lim_{t_\src\to -\infty \atop t_\snk\to\infty}
      e^{E_{q/2}(t_\snk-t_\src)}  \sum_{\vec{x}_\snk, \vec{x}_\src}
      e^{-i\frac{\vec q}{2} \cdot (\vec{x}_\snk + \vec{x}_\src)}
      \label{eq:F_p-src} \\
 &&\hskip 2.5 in \mathcal{F}_\nu(x, y, z, x_\op, x_\snk, x_\src). \nonumber
\end{eqnarray}
Here we will choose the momentum transfer $\vec q = (2 \pi / L)\hat z$, where $\hat z$ is a unit vector in the $z$-direction.  Thus, the muon propagators must be evaluated with anti-periodic boundary conditions in the $z$-direction.  As observed previously, translational symmetry in the three spatial directions and the added factor of $e^{i\vec{q}\cdot\vec{x}_\op}$ introduced in Eqs.~\eqref{eq:M_p-src} and \eqref{eq:F_p-src} imply that the right-hand side of Eq.~\eqref{eq:M_p-src-space} is independent of $\vec{x}_\op$.   Similarly, the right-hand side of Eq.~\eqref{eq:M_p-src-space} does not depend on $t_\op$ since the energies of the initial and final muons are the same.

We can exploit the space-time translational covariance of $\mathcal{F}_\nu(\vec q, x, y, z, x_\op)$ to write the sum in Eq.~\eqref{eq:M_p-src-space} in terms of variables expressed relative to the location of the quark loop.  Begin by shifting all four position arguments of this function by the average $w=(x+y)/2$
\begin{eqnarray}
  \mathcal{M}_\nu(\vec q)  &=& \sum_{x,y,z} e^{i\vec{q}\cdot(\vec{x}_\op-\vec w)} \mathcal{F}_\nu\left(\vec q, \frac{x-y}{2}, -\frac{x-y}{2},  z-w, x_\op-w\right) 
\label{eq:M_p_shifted_sum_1} \\
 &=& \sum_r \left\{ \sum_{\widetilde{z}, \widetilde{x}_\op} e^{i\vec{q}\cdot\vec{\widetilde x}_\op} \mathcal{F}_\nu\left(\vec q, \frac{r}{2},-\frac{r}{2},\widetilde{z},\widetilde{x}_\op\right)\right\},
\label{eq:M_p_shifted_sum_2}
\end{eqnarray}
where in the second equation we have changed summation variables to
\begin{equation}
r = x-y, \quad \widetilde{z} = z -w \mbox{\ \ \ and\ \ \ } \widetilde{x}_\op= x_\op-w
\end{equation}
and explicitly organized the sums so that the sum over the relative coordinate $r$ is performed last.

The form of Eq.~\eqref{eq:M_p_shifted_sum_2} suggests a natural strategy for its evaluation in lattice QCD.   First we make a random choice of the average variable $w$ somewhere within the space-time volume of our simulation.  To match our assumption that $t_\snk-t_\op$ and $t_\op-t_\src$ are large we choose the times $t_\snk$ and $t_\src$ to be $(x_\op)_0+T/4$ and $(x_\op)_0-T/4$ respectively where the sums should be performed modulo $T$, the temporal extent of the lattice volume.  Next the space-time variable $r$ is chosen stochastically as described below and the points $r/2$ and $-r/2$ are used as source locations for two propagators whose sinks are joined at the positions $\widetilde{z}$ and $\widetilde{x}_\op$ which are then explicitly summed over the entire lattice.  The resulting $\mathcal{M}_\nu(\vec q)$ when summed over $w$ and $r$ and averaged over gauge configuration is the desired muon-current 3-point function.

\begin{figure}[!htbp]
  \begin{center}
    \resizebox{0.7\columnwidth}{!}{\includegraphics{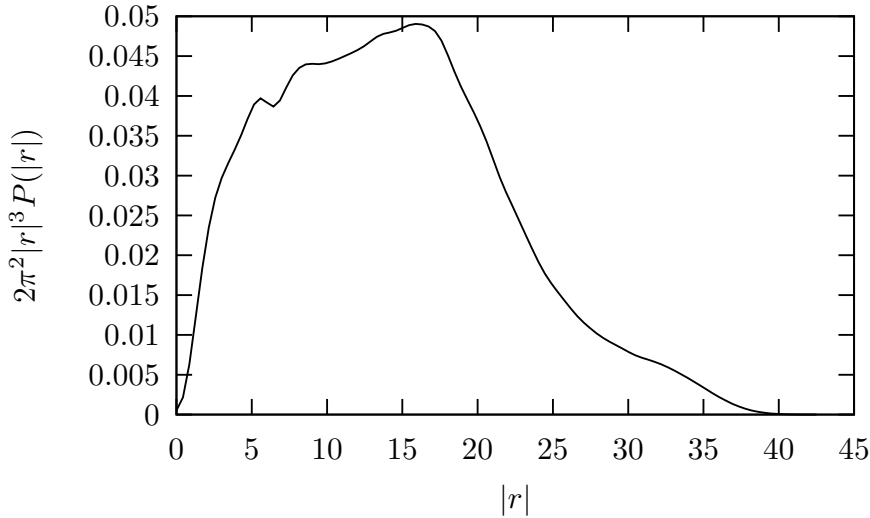}}
  \end{center}
\caption{Distribution of relative separations $|r|=|x-y|$ between the $x$-$y$ pairs of randomly chosen points used to compute $\mathcal{M}_\nu(\vec q)$ on the $32^3\times64$ QCD gauge ensemble described in Sec.~\ref{sec:exact_props_numerical}}.
\label{prob-ceil-r}
\end{figure}

To evaluate the stochastic sum over $r$ efficiently, we use importance sampling, {\it i.e.} we sample most frequently the important region where $|r| \lesssim 1$ fm.   For some of the results presented here we choose a set of $M$ points $\{x_i\}_{1 \le i \le M}$ following the empirically chosen distribution:
\begin{eqnarray}
  p\bigl(|x_i - w|\bigr) & \propto & \left\{ \begin{array}{ll}
    1                         & |x_i - w| < 1\\
    1 / |x_i - w|^{3.5} & |x_i - w| \geqslant 1
  \end{array} \right. ,
\end{eqnarray}
where the special treatment when $|x_i-w|$ is smaller than one lattice unit has been introduced to avoid the singularity in our distribution at $x_i-w=0$.  The distribution of the relative distance $|r|$ between any two points drawn from this set is:
\begin{eqnarray}
  P (|r|) & = & \sum_x p\bigl(|x - r|\bigr) p\bigl(|x|\bigr) .
\end{eqnarray}
The resulting distribution $P(|r|)$ used for our $32^3 \times 64$ ensemble is shown in Fig.~\ref{prob-ceil-r}.

Note, $M(M-1)/2$,  $x$-$y$ pairs can be formed from a set of $M$ points.  (Here each ``pair'' is already symmetrized between the points $x$ and $y$.)   If we calculate a single, point-source quark propagator for each of these $M$ points, then for each $x$-$y$ pair, we can sum over $\widetilde{z}$ and $\widetilde{x}_\op$ exactly with no further inversions.  We find that the resulting statistical error corresponds to that from the larger number of $M^2$ samples unless $M$ is so large that these many samples, all distributed within $\approx 1/m_\pi$ of the single point $w$ become correlated.  This $M^2$ benefit is seen for $M$ at least as large as 16.

In contrast with the stochastic electromagnetic field discussed in Sec.~\ref{sec:stochastic_field}, the statistical noise in the exact photon propagator method remains finite in the infinite volume limit because the quark propagator decreases exponentially with distance.  The noise associated with the stochastic sampling of the space-time points $x$ and $y$ will also fall as $1/\sqrt{N}$ in the limit of a large number $N$ of $x$-$y$ samples provided the distribution $P(|r|)$ that we choose is normalizable in the infinite volume limit, a choice which is certainly possible, again because the quark propagator decreases exponentially with distance.

This exact photon propagator method gives a very large reduction in statistical errors when compared to the previous methods based on a stochastic photon field and is the basis for the $m_\pi=171$ MeV, (4.6 fm$)^3$ volume calculation reported in the next section.  The replacement of a stochastic average over $4 L^3 T$ gauge variables by the simpler importance sampling of two, four-dimensional space-time positions $r$ and $w$ results in a calculation that appears easier to optimize.   We learn {\it a posteriori} how the integrand depends on $|r|$ and can adjust our sampling weights to increase the effectiveness of the sampling.  In particular, we recognize that the largest integrand results from small $|r|$ and therefore compute all pairs with $|r| \leqslant r_{\max}$.  A similar advantage from the use of exact photon propagators may be found when this approach is applied to other processes which include electromagnetism.

\subsection{Current conservation on each configuration}
\label{sec:WI_per_config}

As can be seen from Eq.~\eqref{eq:vertex-func_1}, the form factor $F_2(q^2)$ from which $g_\mu-2$ can be determined is proportional to $q$ which implies that the signal that results from our Monte Carlo average will vanish in the $q\to0$ limit that is needed to determine $F_2(0)$.  However, the form shown in Eq.~\eqref{eq:vertex-func_1} and especially the proportionality to $q$ is a consequence of the conservation of the current $J_\nu$, a condition that will not be obeyed for the individual samples that are averaged in the exact photon propagator method described in the previous section.

As discussed in Appendix~\ref{sec:conserved-vs-local}, if at least one of the four currents coupled to the quark loop is exactly conserved at finite lattice spacing, the HLbL amplitude will be convergent and have a correct continuum limit.  We meet this requirement by using the exactly-conserved, 5-dimensional DWF current as the external current $J_\nu(x_\op)$.  This guarantees that the resulting amplitude will have the form given in Eq.~\eqref{eq:vertex-func_1} up to finite lattice spacing corrections.  However, for the method described in the previous section, the vertices $x$, $x_\op$, $y$ and $z$ appear in a specific order on the quark loop.  We have not computed all three possible insertions for the external photon.  Consequently, the individual samples will not yield a conserved current.  The Ward identity necessary for the external current to have a vanishing divergence will be obeyed only after the stochastic average over $x$ and $y$, which makes the three internal photon vertices on the quark line indistinguishable.  As a result, the noise will not vanish when $q = 0$.

\begin{figure}[!htbp]
  \begin{center}
    \resizebox{0.33\columnwidth}{!}
    {\includegraphics{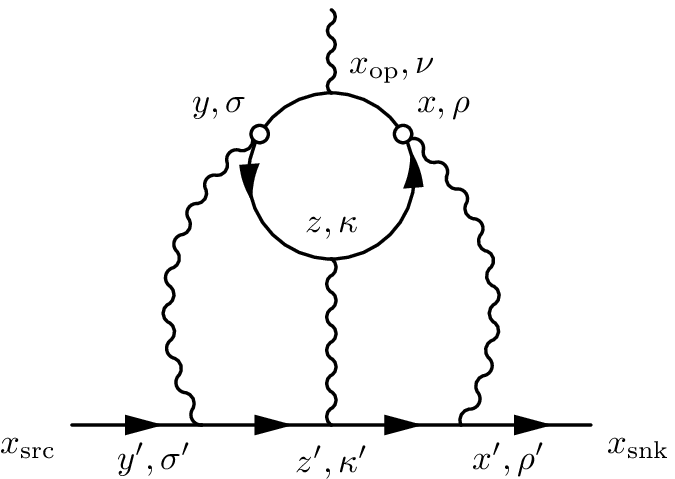}}\resizebox{0.33\columnwidth}{!}
    {\includegraphics{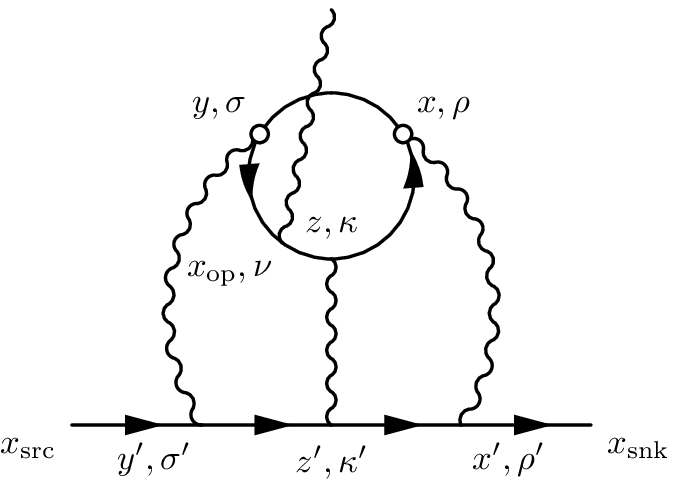}}\resizebox{0.33\columnwidth}{!}
    {\includegraphics{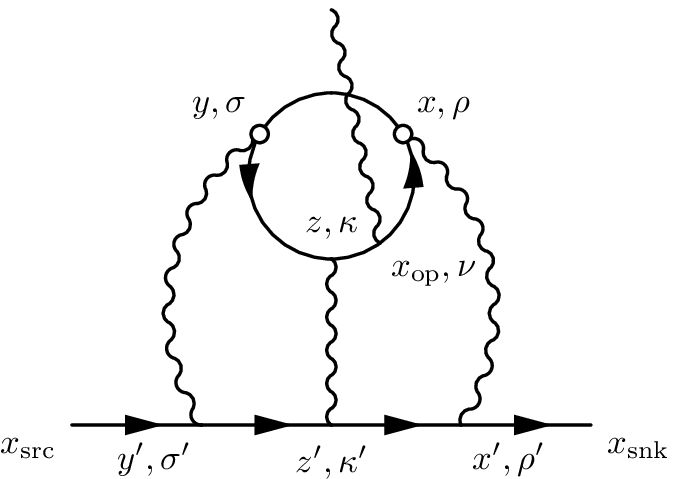}}
  \end{center}
\caption{Diagrams showing the three different possible insertions of the external photon when the vertices $x$ and $y$ are fixed. For each of these three diagrams there are five other possible permutations of the connections between the three internal photons and the muon line that are not shown.  The contributions of each of these three sets of six contractions will be the same after the stochastic average over the vertices $x$ and $y$.}
\label{fig:conserved_diags}
\end{figure}

To make the contribution of each configuration (and hence the statistical noise) vanish as $q\to0$, we must average the three diagrams in Fig.~\ref{fig:conserved_diags} so that the required Ward identity is obeyed, configuration by configuration~\footnote{Although current conservation is exact, in a finite lattice volume with periodic boundary conditions, around the world effects will contribute to the signal and the noise even when the external momentum is zero. However, this noise is suppressed exponentially in the large volume limit. In summary, in the small $q$ and large volume limit, the noise will behave as $\mathcal{O} (q) +\mathcal{O} (e^{- m_\pi L / 2})$.}.   Explicitly, this average can be achieved by replacing the function $\mathcal{F}_\nu$ of Eq.\eqref{eq:hlbl-amp_2} by the symmetrized version $\mathcal{F}^C_{\nu}$ given by:
\begin{eqnarray}
  &  & \mathcal{F}^C_{\nu} \left( x, y, z, x_{\text{op}}, x_{\text{snk}},
  x_{\text{src}} \right) \nonumber\\
  &  & \hspace{1cm} = \frac{1}{3} \mathcal{F}_{\nu} \left( x, y, z,
  x_{\text{op}}, x_{\text{snk}}, x_{\text{src}} \right) + \frac{1}{3}
  \mathcal{F}_{\nu} \left( y, z, x, x_{\text{op}}, x_{\text{snk}},
  x_{\text{src}} \right) \nonumber\\
  &  & \hspace{2cm} + \frac{1}{3} \mathcal{F}_{\nu} \left( z, x, y,
  x_{\text{op}}, x_{\text{snk}}, x_{\text{src}} \right). \label{lbl-amp-c}
\end{eqnarray}
In later equations we will simply add the superscript $C$ to indicate that such an average has been performed.  These additional diagrams are also computationally accessible. The left-hand diagram represents the single amplitude that would be computed following the method of Sec.~\ref{sec:stochastic_field}. The center diagram requires the computation of sequential source propagators at $x_\op$ for each polarization of the external photon.  Finally the right-hand diagram also requires sequential source propagators at $x_\op$, but with the external photon momentum in the opposite direction, since $\gamma_5$-hermiticity must be used to reverse the direction of the propagators, which reverses the momentum of the external photon as well.

Thus, in addition to the point-source propagators from the sites $x$ and $y$, we must compute sequential source propagators as discussed in Sec.~\ref{sec:stochastic_field} for each possible polarization and momentum of the external current. We normally evaluate the amplitude for three polarization directions $x$, $y$, and $t$ (which are perpendicular to the $z$-direction of the external momentum) and two momentum directions (since in some cases the complex conjugate of the sequential source propagator is needed). This requires an additional six times more quark, Dirac-operator inversions. Since we can adjust $M$ to re-balance the cost, the over all cost increase may not be significant but the potential gain can be large, especially in a large volume when we study small $q=2\pi/L$.

There is an additional optimization that can be exploited when all three groups of the diagrams represented in Fig.~\ref{fig:conserved_diags} are computed.  Since the three internal photon vertices are now treated symmetrically, we are free to introduce one asymmetry and restrict the sum over the $z$ vertex to the region where $| x - y | < | x - z |$ and $| x - y | < | y - z |$ and multiply the result by $3$ \footnote{The necessary combinatoric factor is also introduced for the boundary cases where $| x - y | = | x - z |$ or $| x - y | = | y - z |$.}.  This restriction on $z$ will skew the distribution of $|x-y|$ enhancing the region where $| x - y |$ is small and the signal least noisy but suppress the large $|x-y|$ region where the signal is weak and the noise large.

\subsection{Moment method: obtaining $q^2=0$ in finite volume}
\label{sec:moment}

As can be seen in Eq.~\eqref{eq:vertex-func_1}, a matrix element of the current $J_\nu(x_\op)$ between muon states contains the electromagnetic form factor $F_2(q^2)$ multiplied by components of the momentum transfer $q_\rho$. This suggests that $F_2(q^2)$ can be obtained in a lattice calculation only when $q_\beta \ne 0$ so that the anomalous moment $g_\mu-2 = 2 F_2(0)$ can be determined only after taking the limit $q_\rho\to0$.  Of course, this limit is difficult to evaluate in a lattice calculation since the smallest, non-zero momentum component is $2\pi/L$ suggesting that $F_2(0)$ will only become accessible if very large spatial lattice sizes are studied.  We will now show how this potential difficulty can be avoided for the case of the light-by-light contribution to $g_\mu-2$ by evaluating a carefully defined spatial moment of the Feynman amplitude which determines the matrix element of $J_\nu(x)$.

We begin with Eq.~\eqref{eq:M_p_shifted_sum_2} repeated here with a small change in notation:
\begin{eqnarray}
\mathcal{M}_\nu(\vec q)  &=& \sum_{r,z,x_\op}
         \mathcal{F}^C_\nu\Bigl(\vec q, \frac{r}{2},-\frac{r}{2},z,x_\op\Bigr) e^{i\vec q \cdot \vec{x}_\op},
\label{eq:M_p_shifted_sum_3}
\end{eqnarray}
where we have altered that earlier equation by dropping the tilde on the summation variables $z$ and $x_\op$ and adding the superscript $C$. Note that the function $\mathcal{F}^C$ has the same dependence on $x_\op$ as does the current $J_\nu(x_\op)$ whose matrix element is being evaluated and will therefore obey the same Ward identity:
\begin{equation}
\Delta_{(x_\op)_\nu} \mathcal{F}^C_\nu(\vec q, x,y,z,x_\op) = 0,
\label{eq:WI}
\end{equation}
where a sum over the repeated index $\nu$ is understood and $\Delta_x$ evaluates the ``backward''  lattice difference:
\begin{equation}
\Delta_x f(x) = f(x) - f(x-a),
\end{equation}
where $a$ is the lattice spacing.

The critical step in our presentation replaces the factor $e^{i\vec q\cdot \vec x_\op}$ in Eq.~\eqref{eq:M_p_shifted_sum_3} by $(e^{i\vec q\cdot \vec x_\op}-1)$ giving:
\begin{equation}
\mathcal{M}_\nu(\vec q)  = \sum_{r,z,x_\op}
         \mathcal{F}^C_\nu\Bigl(\vec q, \frac{r}{2},-\frac{r}{2},z,x_\op\Bigr)\bigl( e^{i\vec q \cdot \vec{x}_\op}-1\bigr).
\label{eq:M_p_shifted_sum_4}
\end{equation}
The extra `$-1$' term introduced into the sum over $x_\op$ will vanish because of the Ward identity, Eq.~\eqref{eq:WI}, if a surface term can be neglected.  This can be seen from the following manipulation:
\begin{eqnarray}
0 &=& \sum_{x_\op} \Delta_{(x_\op)_\rho} \Biggl( (x_\op)_\nu
    \mathcal{F}^C_\rho\left(\vec q, \frac{r}{2},-\frac{r}{2},z,x_\op\right) \Biggr) \label{eq:cc_1} \\
  &=& \sum_{x_\op} \left\{
     \mathcal{F}^C_\nu \left(\vec q, \frac{r}{2},-\frac{r}{2},z,x_\op\right)
        +(x_\op)_\nu \Delta_{(x_\op)_\rho}   \mathcal{F}^C_\rho\left(\vec q, \frac{r}{2},-\frac{r}{2},z,x_\op\right) \right\} \\
  &=& \sum_{x_\op}  \mathcal{F}^C_\nu \left(\vec q, \frac{r}{2},-\frac{r}{2},z,x_\op\right),
\end{eqnarray}
where the final line demonstrates that the extra `$-1$' term that was added to Eq.~\eqref{eq:M_p_shifted_sum_3} sums to zero.

Finally we can expand the right-hand side of Eq.~\eqref{eq:M_p_shifted_sum_4} in $q_\rho$ and determine
\begin{equation}
\frac{\partial}{\partial q_i} \mathcal{M}_\nu(\vec q)|_{\vec q=0}  = i\sum_{r,z,x_\op}
         \mathcal{F}^C_\nu\Bigl(\vec q=0, r,-r,z,x_\op\Bigr) (x_\op)_i .
\label{eq:M_p_shifted_sum_5}
\end{equation}
While this equation has been derived in infinite space-time volume, the fact that the average of the two points $r$ and $-r$ is located at the origin implies that the integrand decreases exponentially as $|x_\op|$ increases so that this integral can be evaluated in finite volume with only exponentially small corrections.

As discussed in Sec.~\ref{sec:WI_per_config},  Eq.~\eqref{eq:WI} representing current conservation is somewhat subtle.  This equation with the fixed vertices $\pm r$ and $z$ will only be obeyed if the external current $J_\nu(x_\op)$ is inserted in all possible places along the internal quark loop.  This requires that all three diagrams shown in Fig.~\ref{fig:conserved_diags} be included.  This requirement that all three diagrams must be included remains valid even if we perform the integration over the four-vectors $r$ and $z$.   Since the midpoint of the vertices $\pm r$ remains at the origin, these two $\pm r$ vertices remain distinguished and the cancellation required to derive the Ward identity for a closed fermion loop will not be realized unless all three diagrams are combined.

A further refinement of this approach which we have not yet explored numerically, chooses the origin with respect to which $x_\op$ is defined not as the average of the two points $x$ and $y$ but instead as the average of the three points $x$, $y$ and $z$.  With this more symmetrical choice of the origin, the necessary Ward identity would hold when the six possible contractions to the muon line are included and the points $x$ and $y$ stochastically summed.  This approach would then allow us to avoid the calculation of the additional six sequential source propagators that are required when all three diagrams of Fig.~\ref{fig:conserved_diags} must be computed.

We can obtain a complete expression for $F_2(0)$ and hence $g_\mu-2$ from Eq.~\eqref{eq:M_p_shifted_sum_3} by performing a similar small $q$ expansion of Eq.~\eqref{eq:vertex-func_2}.  For the light-by-light diagram in the small momentum transfer limit, we can specialize Eq.~\eqref{eq:vertex-func_2}
\begin{equation}
\Bigl(\frac{-i \slashed q^+ +m_\mu}{2E_{q/2}}\Bigr) \left( \frac{F_2(q^2)}{2 m_\mu} \frac{i}{2}[\gamma_{\nu}, \gamma_{\beta}] q_{\beta} \right) \Bigl(\frac{-i \slashed q^- +m_\mu}{2E_{q/2}}\Bigr) = \mathcal{M}_\nu(\vec q),
\end{equation}
where the external four-momenta $q^\pm = (i E_{q/2}, \pm \vec{q}/2)$. If we examine the case $\nu=i$, equate the coefficients of $(\vec q)_j$, and evaluate the matrix element of this equation between Dirac positive-energy, zero-momentum eigenstates we find
\begin{equation}
\overline{u}(\vec q=\vec 0,s')\left( \frac{F_2(q^2=0)}{2 m_\mu}  \frac{i}{2}[\gamma_i, \gamma_j]\right) u(\vec q=\vec 0,s)
  = \overline{u}(\vec q=\vec 0,s') \frac{\partial}{\partial q_j}\mathcal{M}_i(\vec q)_{\vec q=\vec 0} u(\vec q=\vec 0,s).
\end{equation}
Finally we can multiply the left- and right-hand sides of this equation by $\frac{1}{2}\epsilon_{ijk}$, sum over $i$ and $j$ and use Eq.~\eqref{eq:M_p_shifted_sum_3} to replace the derivative of $\mathcal{M}(\vec q)$ with respect to $q_j$ by the moment of $\mathcal{F}^C$ times $(x_\op)_j$.  The result is the $k^{th}$ component of the vector equation:
\begin{equation}
\frac{F_2(0)}{2 m_\mu}\overline{u}(\vec 0,s')\,\vec\Sigma\,  u(\vec 0,s) =
\frac{1}{2}\sum_{r,z,x_\op} \vec{x}_\op \times i\overline{u}(\vec 0,s')\, \vec{\mathcal{F}}^C \left(\frac{r}{2},-\frac{r}{2},z,x_\op\right) u(\vec 0,s),
\label{eq:moment_Xprod}
\end{equation}
where $\Sigma_i=\frac{1}{4i}\epsilon_{ijk}[\gamma_j, \gamma_k]$.
Here, $i \vec{\mathcal{F}}^C$ represents the quantum-mechanical current and the above equation resembles the conventional expression for the magnetic moment created by a static, localized current~\cite{jackson_classical_1999}:
\begin{equation}
\vec\mu = \frac{1}{2} \int d^3 r \; \bigl[\vec r \times \vec J(\vec r)\bigr].
\label{eq:classical}
\end{equation}
The precise connection between Eqs.~\eqref{eq:moment_Xprod} and \eqref{eq:classical} is worked out in Appendix~\ref{sec:classical_connection}.

\section{Numerical studies} \label{sec:studies}

In this section we describe our numerical results.  This discussion is divided into five subsections.  In the first, subsection~\ref{sec:setup}, we describe the QCD gauge ensembles used in the calculation and explain our treatment of the electromagnetic degrees for freedom, in particular our method for treating the zero or near-zero modes of the photon field in finite volume.  We also explain how the form factor $F_2(q^2)$ is determined from the Euclidean-space correlator that we evaluate.   The second subsection, \ref{sec:examples}, describes a series of example calculations exploring the statistical properties of four techniques that can be used in the calculation of cHLbL using a stochastic representation for the photon propagator described in Section~\ref{sec:stochastic_field}.  In the third subsection, \ref{sec:exact_props_numerical}, we describe in more detail the use of exact photon propagators whose source points are chosen stochastically, giving both our methods and results, including results for the large $4.6 \mathrm{fm}$, $32^3\times 64$ volume and 171 MeV pion mass.  In subsection~\ref{sec:moment_numerical} we extend the exact photon propagator method, now computing the moment as proposed in Section~\ref{sec:moment} and present further large-volume, small quark-mass results, now for $F_2(q^2)$ evaluated at $q^2=0$.  In the final subsection below, \ref{sec:QED_results}, we apply the exact photon propagator and moment methods to the calculation of $(g_\mu-2)$ for the case of the QED light-by-light scattering amplitude, in which the internal loop is a muon instead of a quark, examining the vanishing lattice spacing and large volume limits.  This discussion gives a first indication of the size of the systematic errors associated with finite volume and finite lattice spacing in our results.  It also provides a useful consistency check since we can compare our result with that known from conventional perturbation theory.

\subsection{Computational setup} \label{sec:setup}

We have carried out a series of lattice QED and QCD calculations to both develop the methods described in the previous section and to obtain a result of the cHLbL contribution to $g_\mu-2$ using a relatively light pion in large volume.  We will now provide some of the details of those calculations.  The QCD calculations were performed using four ensembles with the pion masses and lattice volumes listed in Tab.~\ref{tab:confs}.  Although each of the ensembles listed in Tab.~\ref{tab:confs} incorporates 2+1 flavors, with two, degenerate light sea quarks and one physical-mass, strange sea quark, we typically calculate the contribution of a single light quark but multiply by the charge factor $(2 / 3)^4 + (- 1 / 3)^4 = 17 / 81$ to obtain the result expected from a mass-degenerate, up and down quark doublet with charges $+2/3$ and $-1/3$.  Most of our results address only this light quark contribution although for the large-volume, light-pion calculation we also include an explicit, physical strange quark contribution.

\begin{table}[h]
  \begin{center}
    \begin{tabular}{lcccccc}
      Label & size & $L_s$ & $a^{- 1}$(MeV) & $m_\pi$(MeV) & $Z_V$ & Ref\\
      \hline\hline
      16I 	& $16^3 \times 32$ & 16	& $1747$ & $423$ & $0.6998(20)$ & \cite{Allton:2007hx}\\
      24I 	& $24^3 \times 64$ & 16	& $1747$ & $423$ & $0.6998(20)$ & \cite{Allton:2008pn}\\
      24IL 	& $24^3 \times 64$ & 16	& $1747$ & $333$ & $0.6991(17)$ & \cite{Allton:2008pn}\\
      32ID 	& $32^3 \times 64$ & 32	& $1371$ & $171$ & $0.6685(36)$ & \cite{Arthur:2012opa} \\
      \hline
    \end{tabular}
  \end{center}
  \caption{List of ensembles used in our calculations.   Two light and one strange sea quark flavor of domain wall fermions were used when generating these ensembles, where $L_s$ is the length of fifth dimension.  The strange quark mass was chosen close to its physical value.  The values for $Z_V$ are obtained from Tables XLIII and III of Refs.~\cite{Aoki:2010dy} and \cite{Arthur:2012opa} respectively.}
  \label{tab:confs}
\end{table}

The ensembles listed in Tab.~\ref{tab:confs} were obtained using domain wall fermions (DWF)~\cite{Blum:2000kn} and the same DWF Dirac operator was used for the quark loop in cHLbL calculation.  However, for the cHLbL calculations on the 32ID ensemble we used a M\"obius variant~\cite{Brower:2012vk} of the DWF operator that was used to generate the ensemble.  This M\"obius Dirac operator used $L_s=12$ and M\"obius parameters $b+c=32/12$ and $b-c=1$, chosen to ensure that the corresponding M\"obius DWF quark propagator agrees at the few 0.1\% level with the DWF quark propagator used when generating the ensemble.  All of the quark propagators used the five-dimensional mass $M_5=1.8$.

We also use the DWF action for the muon. We compute the muon propagators with the five-dimensional mass $M_5 = 1$ and infinite $L_s$. Since all the muon-photon interactions have been explicitly included in our formulae, the muon propagators are free fermion propagators. To calculate these free propagators, we use Fourier transformations and analytic expressions~\cite{Aoki:2002iq}.   This allows us to exploit the physical properties of DWF with essentially the same computation cost as would be required for fermions without chiral symmetry, {\it e.g.} Wilson fermions.  Because the contribution of those cHLbL subgraphs to $g_\mu-2$ which contain one or more photon-muon vertices will have a negative degree of divergence, we can use local currents for the photon-muon coupling at $x'$, $y'$, and $z'$ and incur only $O(a^2)$ errors.

As is discussed in detail in Appendix~\ref{sec:conserved-vs-local}, we can avoid a divergent contact term resulting from the quark loop in the cHLbL diagram if only one of the four vertices where a photon attaches to that quark loop is given by a conserved current.  Thus, we use the complete, five-dimensional, non-local conserved form for the external current while for the three vertices $x$, $y$, and $z$ attached to internal photons we use the simpler local, four-dimensional current in the above formulae.  We introduce the factor of $Z_V^3$ that is needed to properly normalize these three local, non-conserved currents.  (The additional convergence provided by the first, position-space moment of the cHLbL amplitude allows us to use only local currents for that case.)

We use Feynman gauge for the photon propagator which can be written as
\begin{eqnarray}
  G_{{\mu}, \nu} (x, y) & = & \frac{1}{VT}
  \sum_{k \atop |\vec k | \neq 0}
  \frac{\delta_{\mu, \nu}}{\widetilde{k}\,^2} \exp\bigl(ik \cdot
  (x - y)\bigr) ,
  \label{eq:photon_prop}
\end{eqnarray}
where $VT$ is the space-time volume in lattice units.  The four-vector $k=(k_0, \vec k)$ is determined by four integers $k = 2\pi(n_0/T, \vec n/L)$, where the integers $n_\nu$, $0 \le \nu \le 3$ obey $-T/2 < n_0 \le T/2$ and $-L/2 < n_i \le L/2$ for $ 1 \le i \le 3$.  The four-vector $\widetilde{k}$ appearing in the denominator of Eq.~\eqref{eq:photon_prop} is given by
\begin{eqnarray}
  \widetilde{k}_{\nu} & = & 2 \sin \left( \frac{k_\nu}{2} \right).
\end{eqnarray}
The omission of all Fourier modes with $\vec k = 0$ from the sum appearing in Eq.~\eqref{eq:photon_prop} removes a possible infrared singularity and will contribute to the finite-volume error that is present in our results~\cite{Hayakawa:2008an}.

As a first step of generating stochastic photon fields, we define a complex photon field
\begin{eqnarray}
  \mathcal{A}_\nu(x) & = & \frac{1}{\sqrt{VT}}
  \sum_{k\atop| \vec k | \neq 0} \frac{\epsilon_\nu(k)}{\sqrt{(\widetilde{k})^2}}
  \exp(i k \cdot x),
\end{eqnarray}
where $\epsilon_\nu (k)$ is a random complex variable which satisfies
\begin{eqnarray}
  \langle \epsilon_{{\mu}} (k) \epsilon_{\nu}^{\ast} (k') \rangle_A & = &
  \delta_{{\mu}, \nu} \delta_{k, k'},\\
  \langle \epsilon_{{\mu}} (k) \epsilon_{\nu} (k') \rangle_A & = & 0
\end{eqnarray}
and the average $\langle\ldots\rangle_A$ indicates an average over the random variables $\epsilon_\nu (k)$.  In our calculations, we choose $\epsilon_\nu(k)$ to be a Gaussian random
variable, which is similar to the distribution of the gauge fields found in conventional QED gauge ensembles. We can verify that this complex stochastic field will generate the desired Feynman-gauge, photon propagator:
\begin{eqnarray}
  \langle \mathcal{A}_{{\mu}} (x) \mathcal{A}_{\nu}^{\ast} (y) \rangle_A
  &=& \frac{1}{VT} \sum_{k\atop |\vec k| \neq 0} \sum_{k'\atop|\vec k\,' | \neq 0}
  \Bigl\langle \epsilon_\mu (k) \epsilon_\nu^{\ast} (k')
  \Bigr\rangle_A  \frac{e^{   i k \cdot x}}{\sqrt{(\widetilde{k})^2}}
  \frac{ e^{- i k'\cdot y}}{\sqrt{(\widetilde{k}')^2}}\\
  &=& G_{{\mu}, \nu} (x, y) .
\end{eqnarray}
Finally a real stochastic photon field can be constructed from the real part of $\mathcal{A}_{\mu}(x)$:
\begin{eqnarray}
  A_{{\mu}} (x) & = & \sqrt{2}
  \ensuremath{\operatorname{Re}}\mathcal{A}_{{\mu}} (x),
\end{eqnarray}
which obeys:
\begin{eqnarray}
  \langle A_{{\mu}} (x) A_{\nu} (y) \rangle_A & = & \frac{1}{2} (\langle
  \mathcal{A}_{{\mu}} (x) \mathcal{A}_{\nu}^{\ast} (y) \rangle_A + \langle
  \mathcal{A}_{{\mu}}^{\ast} (x) \mathcal{A}_{\nu} (y) \rangle_A)
  =  G_{{\mu}, \nu} (x, y).
\end{eqnarray}
It is this real stochastic photon field $A_\mu(x)$ that we use in the calculation.

\begin{figure}[H]
  \begin{center}
    \resizebox{0.45\columnwidth}{!}{\includegraphics{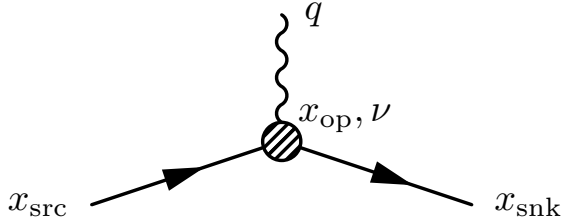}}
  \end{center}
  \caption{The shaded circle represents the vertex: $F_1 (q^2) \gamma_{\nu} + i \frac{F_2 (q^2)}{4 m_\mu}[\gamma_{\nu}, \gamma_{\beta}] q_{\beta}$.}
\label{fig:tree-f1f2-diagram}
\end{figure}

While the three-momenta of the initial and final muons are typically fixed to be $\pm\vec q/2$, we calculate all 16 amplitudes corresponding to all possible initial and final spinor indices, $\alpha$ and $\beta$.  We extract the form factor $F_2(q^2)$ from the resulting $4\times4$ matrices $\mathcal{M}_\nu(\vec q)_{\alpha\beta}$ for different external photon polarizations $\nu$ of Eq.~\eqref{eq:M_p-src} by matching to the Green's function shown in Fig.~\ref{fig:tree-f1f2-diagram}.  This diagram represents the result of a tree-level calculation with a muon source and sink identical to those used in our lattice calculation but with a vertex function that is expressed in terms of the general invariant functions $F_1(q^2)$ and $F_2(q^2)$ using Eq.~\eqref{eq:vertex-func_1}.

We compute the tree-level amplitude $(\mathcal{M}^\text{tree}_\nu(\vec q) )_{\alpha\beta}$ described by the diagram given in Fig.~\ref{fig:tree-f1f2-diagram} as a function of the input variables $F_1(q^2)$ and $F_2(q^2)$ on the same lattice volume, with same muon source and sink momenta as were used in the cHLbL calculation obtaining
\begin{eqnarray}
\mathcal{M}^\text{tree}_\nu(\vec q)  &=&
e^{E_{q/2}(t_\snk-t_\src)}
e^{i\vec{q}\cdot\vec{x}_\op}
\sum_{\vec{x}_\snk, \vec{x}_\src}
e^{-i\frac{\vec q}{2} \cdot (\vec{x}_\snk + \vec{x}_\src)} \\
& & \hskip 0.75in \cdot
S_\mu(x_\text{snk},x_\text{op})
\Big(
F_1 (q^2) \gamma_{\nu} + i \frac{F_2 (q^2)}{4 m_\mu}
[\gamma_{\nu}, \gamma_{\beta}] q_{\beta}
\Big)
S_\mu(x_\text{op},x_\text{src}). \nonumber
\label{eq:M_p-src-tree}
\end{eqnarray}
We then find the two values of $F_1(q^2)$ and $F_2(q^2)$ that minimize the difference
\begin{eqnarray}
\sum_{\nu}
\sum_{\alpha,\beta}
\Biggl|
\Bigl(\mathcal{M}_\nu(\vec q)\Bigr)_{\alpha\beta}
- \Bigl(\mathcal{M}^\text{tree}_\nu(\vec q)\Bigr)_{\alpha\beta}
\Biggr|^2.
\label{eq:tree-difference}
\end{eqnarray}
In most of our simulations, we choose $\vec q$ to be in the $z$ direction.
Since $(\mathcal{M}^\text{tree}_z(\vec q) )_{\alpha\beta}$ will then naturally be zero,
we omit that direction from the above summation.

In the moment method, both muons carry zero momentum and the resulting, simplified spinor structure is given in Eq.~\eqref{eq:moment_Xprod}.  Because $F_1(0)=0$ in this case, we only need to find the minimum with respect to $F_2(0)$ and can neglect the amplitude $(\mathcal{M}_t(\vec q) )_{\alpha\beta}$ corresponding to polarization in the $t$ direction.
However, we still evaluate the tree diagram of Fig.~\ref{fig:tree-f1f2-diagram} and minimize the expression in Eq.~\eqref{eq:tree-difference} to obtain $F_2(0)$.

When computing quark propagators on configurations belonging to the four ensembles described in Tab.~\ref{tab:confs} we use low mode deflation with 100 eigenvectors for the 16I ensemble and 550 eigenvectors for the other three ensembles.  These low modes are also used when computing the reduced-precision propagators that are used in the all-mode-averaging procedure described below.  Except for these low-precision inversions, the Dirac operator was inverted using a stopping condition of $10^{-8}$.  More specifically, we required that the inverse of a product of the preconditioned Dirac operator times its hermitian conjugate solve the Dirac equation with a residual whose norm was $10^{-8}$ times smaller than the norm of the vector to which the inverse was applied.

We conclude this subsection with a discussion of the unconventional strategy which we have implemented in all of numerical work presented here.  In contrast to most lattice QCD calculations the initial and final states do not contain quarks and enter a computationally inexpensive portion of the calculation.  The bulk of the computational effort is associated with evaluating quark propagators whose sources have fixed positions in space-time and are necessarily located close to the position $x_\op$ of the external current $J_\nu$.   In order to suppress the contribution of excited states (typically states of a muon with one or more photons) we must work with large time separations $t_f-t_\op$ and $t_\op - t_i$.  To the extent that these separations are large, our final Green's functions will depend on $t_f$ and $t_i$ only through their difference $t_f-t_i$ which we hold fixed at $T/2$.  In order to achieve the greatest suppression of excited states we will choose the locations of the muon source and sink to be maximally distant from the sources of the quark propagators.  Specifically we locate $t_f$ and $t_i$ so that the average $w=(x+y)/2$ appearing in Eq.~\eqref{eq:M_p_shifted_sum_1} lies midway between $t_f$ and $t_i$.  This means that we do not keep $t_f-t_\op$ and $t_\op - t_i$ fixed but instead average over a range of large values of $t_f-t_\op$ and $t_\op - t_i$, upon which the quantity we are computing should not depend.  In order to provide numerical evidence that the effects of excited states have been reduced below the level of our statistical errors we simply vary $t_{\mathrm{sep}}=t_f-t_i$ to explore the degree to which our results depend on it.

\subsection{Example stochastic photon calculations} \label{sec:examples}

In Section~\ref{sec:stochastic_field} we compared the original subtraction method used to obtain the first lattice QCD results for $g_\mu-2$~\cite{Blum:2014oka} and an alternative stochastic method in which specific random photon fields are introduced to construct only the three propagators needed for the $O(\alpha^3)$ cHLbL amplitudes.  In this section we will not attempt a numerical comparison of these two methods since the absence of both $O(\alpha^2)$ noise and the need to remove unwanted $O(\alpha^4)$ and higher order terms gives the latter method a clear advantage.  (A comparison of the original method and a combination of many of the improvements suggested in this paper can be found in Fig.~\ref{fig:old_new_compare}, presented later in this section.)  Instead we will begin by comparing a series of variations of the stochastic field method.

\label{sec:M-squared}

\begin{table}[ht]
  \begin{center}
    \begin{tabular}{lccccc}
      \hline
Method	&$F_2/(\alpha /\pi)^3$	& $N_{\mathrm{prop}}$ &
      $\sqrt{\text{Var}}$ & $N_{\mathrm{sample}}$ & $\sqrt{\text{Var}_{\text{Eff}}}$\\
      \hline \hline
      Stoch.               & $0.2228(46)$ & $9,864 \times (2 \times (1 + 12))$ & $2.3$  & $9,864 \times 12^2$ & $5.5$ \\
      Stoch. w/o $M^2$ & $0.1962(368)$ & $18,432  \times (2 \times (1 + 1))$ & $10.0$ & $18,432 \times 1^2$ & $5.0$ \\
      Stoch. ext. pt. & $0.232(33) $ & $18,096 \times (2 \times (1 + 6))$ & $16.6$ & $18,096 \times 6^2$ & $28.4$\\
      \hline
    \end{tabular}
  \end{center}
\caption{Results for $F_2$ evaluated at $q^2=(2\pi/L)^2$ for three stochastic propagator methods.  The calculations were performed on a $16^3 \times 64$ lattice with a muon mass of 0.02, a time separation of 32 between the muon source and sink and using an internal muon loop.   For this test we used a local current for the external photon and conserved currents for internal photons.  However, the 2- and 3-photon contact terms needed for these conserved currents were not included.   A summary of these results has been presented in Ref.~\cite{Jin:2014cea}.}
\label{tab:16-qed-0.2}
\end{table}

\subsubsection{$M^2$ method}
\label{sec:M_2-method}

We first study the statistical advantage that results if we compute $M$ sequential source propagators for the $x$ vertex in Fig.~\ref{fig:hlbl_AB} with momentum $-q/2$ injected at the external current vertex and an additional  $M$ sequential source propagators for the $y$ vertex with the momentum $+q/2$ injected and then evaluate all $M^2$ possible pairs.  This test is carried out on a  $16^3\times64$ lattice and uses muon propagators for both the external muon and the internal loop.  Thus, there are no fluctuating QCD configurations and the resulting statistical noise comes entirely from the stochastic photon propagators.

The advantage of using the $M^2$ method can be seen by comparing the first two rows of Table~\ref{tab:16-qed-0.2}.  The first row evaluates $M=12$ stochastic propagators for each of the two sequential sources created from propagators whose sources correspond to the external current with four-momenta $+q/2$ and $-q/2$ and combines them using all $M^2$ possible pairs.  The second row uses two stochastic sequential source propagators corresponding to single sequential sources at the $x$ and $y$ vertices in Fig.~\ref{fig:hlbl_AB}, again carrying the momenta $\pm q/2$.  Both the first and second rows of Table~\ref{tab:16-qed-0.2} use a random space-time volume source for the external current.
The quantity $N_{\mathrm{prop}}$ listed in the third column in Table~\ref{tab:16-qed-0.2} is the total number of propagator inversions required for each result and is given by
\begin{equation}
N_{\mathrm{prop}} =  N_{\mathrm{ext \mathchar`- cur}} \bigl(2(1 + M)\bigr).
\end{equation}
Here the factor $1+M$ corresponds to $1$ random wall (or point) source inversion and $M$ sequential source inversions for $M$ different stochastic photon fields while $N_{\mathrm{ext\mathchar`-cur}}$ is the number of random wall or point sources used for the external current.  The extra factor of ``2'' is needed because the external photon carries momentum which requires two separate momenta for the fermions entering and exiting at this vertex.

As can be seen from the first two rows of Table~\ref{tab:16-qed-0.2}, we realize a substantial reduction in statistical error when using the $M^2$ method.  Since the computational cost involved in these two rows is not the same, a precise comparison requires more than a simple comparison of the resulting statistical errors.   This comparison is assisted by the $\sqrt{\text{Var}}$ and $\sqrt{\text{Var}_{\text{Eff}}}$ columns in that table.  In each of these columns we begin with the quoted jackknife statistical error and compute a measure of the width of distribution of individual samples before the average over samples is performed.  For the quantity $\sqrt{\text{Var}}=\text{Err} \sqrt{N_{\mathrm{prop}}}$  we simply expand the final error ($\text{Err}$) by a factor given by the square root of the number of internal loop propagators that were computed to produce that error.  The comparison of $\sqrt{\text{Var}}$ between the first and second rows suggests that the statistical fluctuations found in the result for a given computational cost were roughly five times smaller for the $M^2$ method.  The quantity $\sqrt{\text{Var}_{\text{Eff}}} = \text{Err} \sqrt{N_{\mathrm{sample}}}$, where $N_{\mathrm{sample}} =  N_{\mathrm{ext\mathchar`-cur}} M^2$ inflates the final quoted error by the square root of the number of ``effective'' samples $N_{\mathrm{sample}}$ which in this case treats the $M^2$ samples as if they were all independent.  Here the resulting nearly equal ``effective'' variances imply that this hypothesis is true and these $M^2$ samples are essentially independent.  Thus, if only the cost of the internal muon line is considered, in this case the $M^2$ method has reduced the computational cost by a factor of $M=12$!

The choice of $M=12$ made in this test was motivated by the case of QCD with an internal quark loop.  In that case the $M^2$ method allows $M^2$ samples from $2M$ computationally expensive light quark propagator inversions.  However, we need to evaluate the product of external muon propagators for all six different permutations of the three internal photons, each pair of stochastic photons joined to $x$ and $y$ and all combinations of photon polarizations.  Since this muonic part of the calculation grows at $M^2$ we cannot make $M$ too large.  In our simulations, the choice $M = 12$  balances the cost of muons and quarks but is not so large that the QCD gauge noise seen from configuration to configuration dominates the statistical noise, so the the statistical gain is still proportion to $M^2$.

\subsubsection{Random wall sources for the external current}
\label{sec:random_wall}

A second, standard method to increase the efficiency of this cHLbL calculation attempts to increase the degree of volume averaging by using a random wall source for the two sequential source propagators appearing in the internal loop, instead of choosing one or more point sources.  For a random source at a given time $t_\op$ we use a full spatial vector of Gaussian random numbers with a different vector being chosen for each spin and color.  As described above, two independent noise vectors are needed with momentum factors $\exp{(\pm i q\cdot x_\op/2)}$.  If the propagators corresponding to one of these two noise sources is multiplied by the complex conjuage of the other which are then combined with the second noise vector and the complex conjugate of the first, an expression can be constructed whose noise average will be the desired sum over all locations of $\vec x_\op$ for a fixed choice of $t_\op$.   Such random volume sources were used to obtain the results given in the first two rows of Tab.~\ref{tab:16-qed-0.2}.

In order to determine the value of this use of a random wall source, we generated the results in the third row of Table~\ref{tab:16-qed-0.2} by using $N_{\mathrm{ext\mathchar`-cur}}$ point source locations for the external current.   (Here the extra factor of two in cost for external current sources carrying the momenta $\pm q/2$ could have been avoided but this would not have changed the qualitative conclusion.)  By comparing the first and third rows of Table~\ref{tab:16-qed-0.2} one sees a 5 to 8 times reduction in the errors from the use of a random wall source.

\subsubsection{Breit-frame muon momenta}
\label{sec:Breit}

The symmetrical choice of $\pm \pi/L$ for the outgoing and incoming momentum has aesthetic appeal and only non-zero spatial momenta as is required for a direct lattice measurement of a magnetic moment.  However, by avoiding assigning a $2\times$ larger spatial momentum of $2\pi/L$ this approach also results in substantially smaller statistical errors than the simpler assignment of the allowed, finite-volume four-momenta $(m_\mu, \vec 0)$ and $(\sqrt{m_\mu^2 + (2\pi/L)^2}, 2\pi/L \hat e)$ to the incoming and outgoing muon momenta.  The errors obtained using this standard assignment and those resulting from the Breit- or brick-wall-frame choice made here, with the incoming and outgoing four-momenta $(\sqrt{m_\mu^2 + (\pi/L)^2}, \mp\pi/L \hat e)$ are compared in Tab.~\ref{tab:16-qed-0.1}.  (Here $\hat e$ is a unit vector in the direction of one of the three spatial axes.)  This comparison is identical to that shown in Tab.~\ref{tab:16-qed-0.2} except the muon mass has been reduced from 0.02 to 0.01 and shows an approximate 15 times reduction in error which is equivalent to what would be obtained with two-hundred times that statistics.  Such a reduction in error should be expected.   When the initial and final momenta are $\vec q$ and $\vec 0$, the signal behaves as $\exp(-(E_{\vec q} + E_{\vec 0}) t_\sep / 2)$.  However, the noise behaves as $\exp(-E_{\vec 0} t_\sep)$, which leads to an exponentially decreasing signal to noise ratio in the large time separation limit.

\begin{table}[H]
\begin{center}
\begin{tabular}{lccccc}
\hline
Method	&$F_2/(\alpha /\pi)^3$	& $N_{\mathrm{prop}}$ & $\sqrt{\text{Var}}$ & $N_{\mathrm{sample}}$ & $\sqrt{\text{Var}_{\text{Eff}}}$\\
\hline \hline
Stoch.  $\vec{p}_1 = -\frac{\pi}{L}\hat e$
					 	& $0.1666(69) $ & $1584 \times (2 \times (1 + 12))$ & $1.4$& $1584  \times 12^2$ & $3.3$ \\
Stoch. $\vec{p}_1 = \vec 0 $	& $0.2278(265) $ & $10260 \times (2 \times (1 + 24))$ & $19.0$ & $10260  \times 24^2$ & $64.4$ \\
\hline
\end{tabular}
\end{center}
\caption{Comparison of results obtained with muon momenta of $\pm q/2 \hat e$ using twisted boundary conditions for the initial and final muon propagators with those obtained when the initial muon carries zero momentum and the final muon is given $q \hat e$.  Here $q=2\pi/L$, and $\hat e$ is a unit vector parallel to one of the edges of the spatial volume.  Except for the choice of muon mass, $m_\mu = 0.01$, all features of the calculation and definitions are the same as those for Table~\ref{tab:16-qed-0.2}.  A summary of these results has been presented in Ref.~\cite{Jin:2014cea}.}
\label{tab:16-qed-0.1}
\end{table}

\subsection{Exact photon propagators}
\label{sec:exact_props_numerical}

The use of exact instead of stochastic photon propagators is the most significant improvement in method suggested in this paper because of its elimination of stochastic noise which grows with the volume.  In this subsection we describe the implementation of this method, compare it with our earlier results and apply it to obtain the cHLbL contribution to $g_\mu-2$ for near-physical circumstance with $m_\pi=171$ MeV and a reasonably large, $32^3\times 64$ lattice volume which is 4.6 fm on a side in physical units.

As described earlier, we choose stochastically the location of two of the three vertices $x$ and $y$ at which the internal photons couple to the quark loop.  The pair of positions $x$ and $y$ are point sources for the quark propagators and we arrange the contractions so that the location of the external current, $x_\op$, and the third photon vertex, $z$ appear as sinks and are explicitly summed over space-time.  While computational cost prevents our performing an explicit sum over all space-time separations $r_\nu=x_\nu-y_\nu$, we can split the computation of the sum into two parts. The first part contains all $r_\nu$ values with Euclidean  magnitude less than a certain value: $|r| \le r_{\max}$.  Here we evaluate all distinct separations $r_\nu$ up to discrete symmetries.  The second part of the sum, where the magnitude $|r|$ is larger than $r_{\max}$, is evaluated by averaging over random point-pair samples, weighted to increase the sampling efficiency.

\begin{table}[H]
  \begin{center}
    \begin{tabular}{lcccc}
      \hline
      Method & $F_2/(\alpha / \pi)^3$ & $N_{\mathrm{confs}}$ &
      $N_{\mathrm{prop}}$ & $\sqrt{\text{Var}}$\\
      \hline
      \hline
      Stoch. & $0.1485(116)$ & 31 & $32 \times (2 \times (1 + 6))$ &
      $1.37$\\
      Exact & $0.1235(26)$ & 16 & $129 + 16 \times 16$ & $0.051$\\
      \hline
    \end{tabular}
  \end{center}
\caption{Comparison of the stochastic and exact photon methods carried out on the 16I ensemble with $m_{\mu} = 332 \mathrm{MeV}$ and the separation between the muon source and sink $t_{\text{sep}} = 16$.  As in the previous tables $\sqrt{\text{Var}} =\text{Err} \sqrt{N_{\mathrm{confs}} N_{\mathrm{prop}}}$.  Here $N_{\mathrm{confs}}$ is the number of configurations analyzed and $N_{\mathrm{props}}$ the number of propagators that are computed on each configuration.  In both cases $F_2(q^2)$ is evaluated at the minimum, non-zero lattice momentum transfer $(2\pi/L)^2$.}
\label{tab:16I-cmp}
\end{table}

We compare the exact-photon method with our previous stochastic method by performing a test on the 16I ensemble.  The results are listed in Tab.~\ref{tab:16I-cmp}.  For the stochastic method, the total number of propagator inversions per configuration, $N_{\mathrm{props}} = N_{\mathrm{set}} \bigl(2(1 + M)\bigr)$ where $1+M$ corresponds to 1 random wall source inversion and $M$ sequential source inversions for $M$ different stochastic photon fields.  The quantity $N_{\mathrm{set}}$ is the product of the number of random sources used per time slice and the number of time slices used on each configuration analyzed.  For the exact-photon method, $N_{\mathrm{props}} = N_{\mathrm{short\mathchar`-dist}} +N_{\mathrm{set}} M$.  In the ``stochastic'' method, we use a local current for the external photon and the conserved current for the internal photons, with the necessary contact terms included in these cases.  In the ``exact-photon'' method, we use the conserved current for the external photon and a local current for each internal photon coupling. (There are no contact terms required in this case.)  We can see that even on this relatively small volume the exact-photon method is more than 700 times as cost effective as the stochastic method.

\begin{table}
  \begin{center}
    \begin{tabular}{lccc}
      \hline
      Label & $m_{{\mu}} /\mathrm{MeV}$ & $N_{\mathrm{confs}}$ &
      $F_2/(\alpha / \pi)^3$\\
      \hline
      \hline
      16I & $332$ & $16$ & $0.1235(26)$\\
      24I & $332$ & $17$ & $0.2186(83)$\\
      24IL & $261$ & $18$ & $0.1570(69)$\\
      32ID & $134$ & $47$ & $0.0693(218)$\\
      32ID-S & $134$ & 23 & $0.0195(88)$\\
      \hline
      Model &  &  & $0.08(2)$\\
      $\mathrm{Exp}-\mathrm{SM}$ &  &  & $0.28(7)$\\
      \hline
    \end{tabular}
  \end{center}
  \caption{The magnetic form factor $F_2 (q^2)$ evaluated at $q^2 = (2 \pi / L)^2$ for our four ensembles. In each case, we choose the muon mass to give the physical value for ratio of muon to pion mass.  The 32ID-S results are obtained from the 32ID ensemble but with the loop mass set to that of the strange instead of the light quark. The actual strange quark contribution to cHLbL for the 32ID ensemble would be the value shown divided by $17$ to introduce the proper electric charge weighting. The last two lines are for comparison: ``Model'' is the result presented at the Glasgow meeting~\cite{Prades:2009tw} and `` $\mathrm{Exp}-\mathrm{SM}$'' is the E821 experimental value minus the standard model prediction, without a HLbL contribution.}
\label{tab:psrc-results}
\end{table}

The results for $F_2(q^2)$ at $q^2=(2\pi/L)^2$ using the exact-photon method for each of the ensembles listed in Tab.~\ref{tab:confs} are presented in Tab.~\ref{tab:psrc-results}.  The statistical weights for the separations between the pairs and other simulation parameters used to obtain these results are listed in Tab.~\ref{tab:psrc-qcd-sims-params}.

\begin{table}
  \begin{center}
    \begin{tabular}{lccrrc}
      \hline
      Label & $r_{\max}$ & $p (x)$ & $M$ & $N_{\mathrm{set}}$ & $\frac{\text{Cost per
  conf}}{\text{BG/Q rack days}}$\\
  \hline\hline
  16I & $4$ & $1 / |x|^{3.5}$ & $16$ & 16 & $0.039$\\
  24I & $4$ & $1 / |x|^{3.5}$ & $16$ & 16 & $0.178$\\
  24IL & $4$ & $1 / |x|^{3.5}$ & $16$ & 16 & $0.177$\\
  32ID & $3$ & $1 / |x|^{3.5}$ & $16$ & 8 & $0.224$\\
  32ID-S & $4$ & $1/|x|^4$ & $8$ & 8 & $0.085$\\
  \hline
\end{tabular}
  \end{center}
\caption{Simulation parameters used to obtain the results given in Tab.~\ref{tab:psrc-results}. The quantity $r_{\max}$ is the upper bound on the magnitude of the $x-y$ separations which are evaluated without random sampling, $M$ is the number of randomly sampled points that are combined using the $M^2$ method, while $N_{\mathrm{set}}$ is the number of groups of these $M$ samples analyzed per QCD configuration.  Note, for each set of $M$ random points we randomly chose a point $s$ in the lattice volume and then the $M$ stochastic points which will be used for the vertices $x$ and $y$ are chosen relative to that random point $s$ following the weight $p(x-s)$.}
\label{tab:psrc-qcd-sims-params}
\end{table}

Since we calculate the contribution for each $x$-$y$ pair, the results contain more information than a single final number. In Fig.~\ref{fig:f2-ceil-r}, we plot a histogram of the contributions to $F_2$ from different point-pair separations and a scatter plot of the $F_2$ contribution from each random point-pair sample.  Shown are results for the four different QCD ensembles described in Tab.~\ref{tab:confs}.  The fifth row labeled 32ID-S uses the strange instead of the light quark in the quark loop, evaluated on the 32ID ensemble.   Tab.~\ref{tab:psrc-qcd-sims-params} lists the choices made in sampling the points $x$ and $y$ for each case.  As can be seen in Fig.~\ref{fig:f2-ceil-r}, the majority of the contribution to $F_2$ comes from a separation of $|r| \le 10$ in lattice units or $|r| \le 1.4~\mathrm{fm}$.  However, most of the statistical noise comes from the more difficult to sample, larger separations with $|r| \ge 1.4~\mathrm{fm}$, even for the case of the heavier strange quark.

\begin{figure}
  \begin{center}
    \resizebox{0.38\columnwidth}{!}{\includegraphics{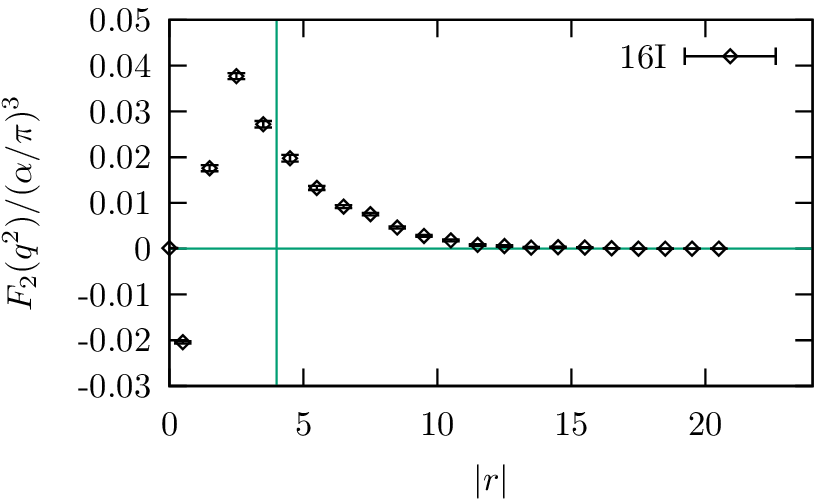}}\resizebox{0.38\columnwidth}{!}{\includegraphics{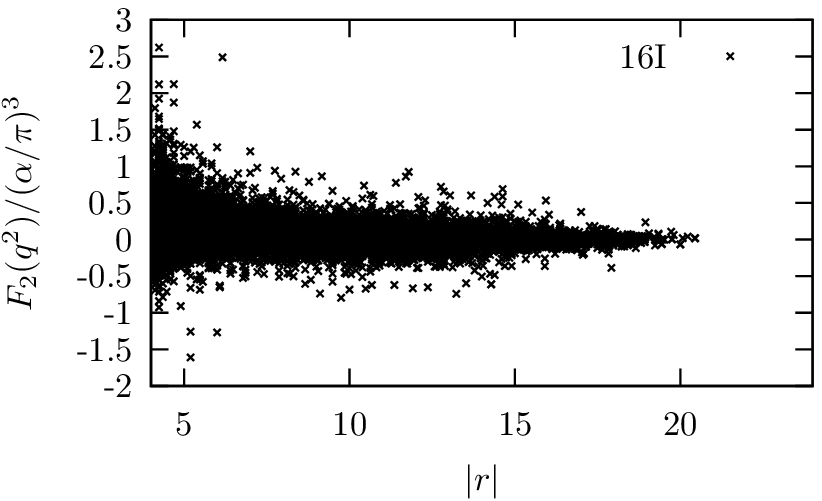}}
\\
    \resizebox{0.38\columnwidth}{!}{\includegraphics{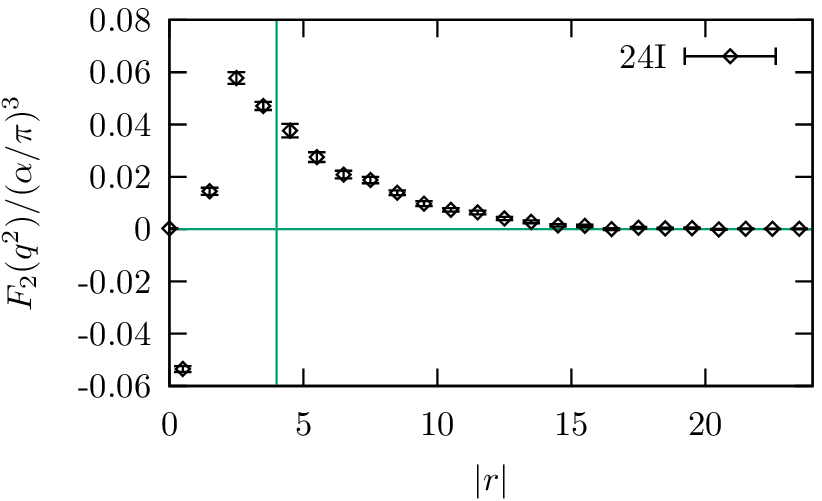}}\resizebox{0.38\columnwidth}{!}{\includegraphics{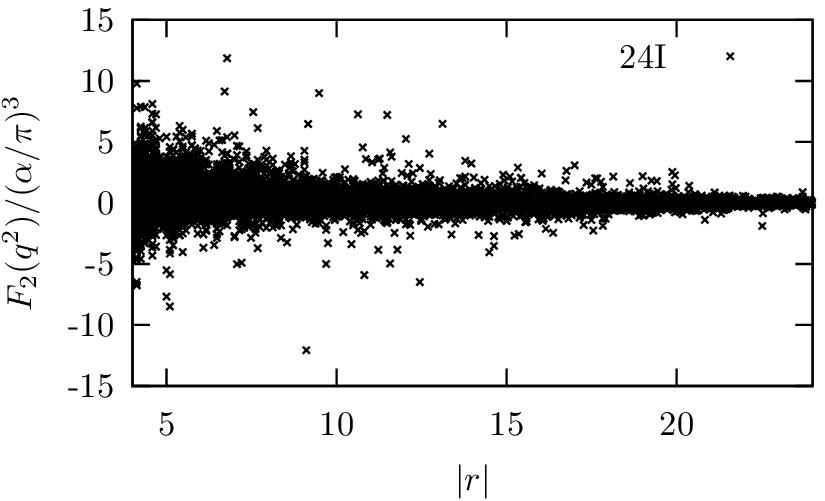}}
\\
    \resizebox{0.38\columnwidth}{!}{\includegraphics{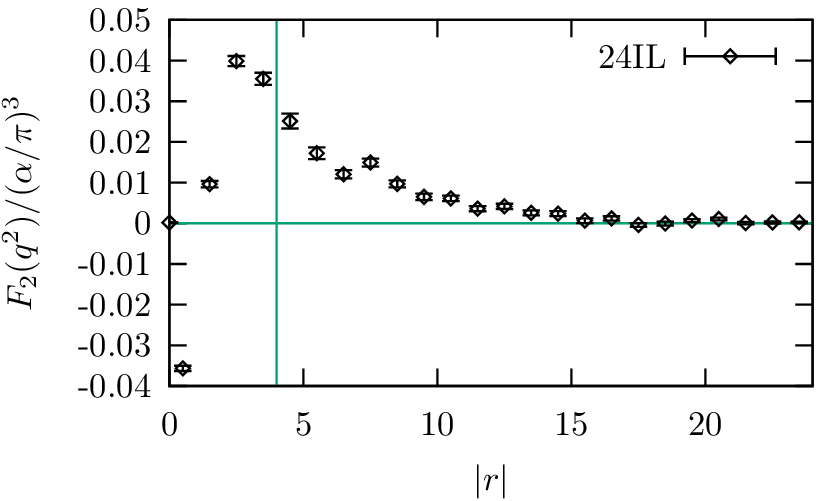}}\resizebox{0.38\columnwidth}{!}{\includegraphics{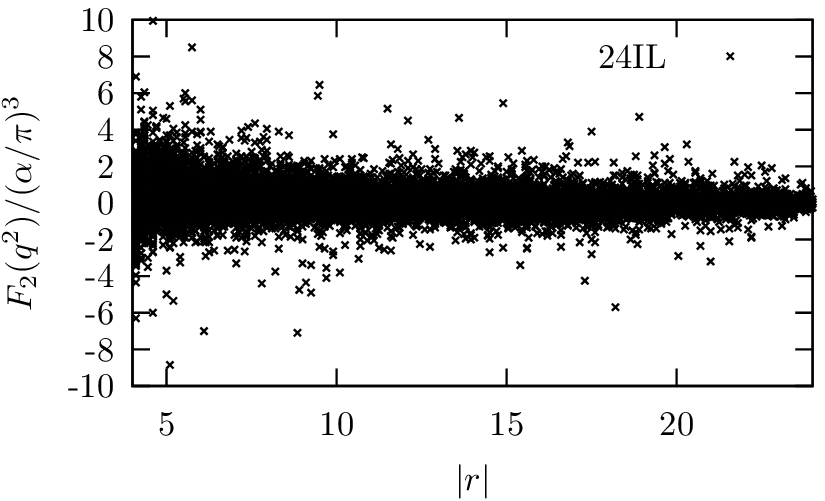}}
\\
    \resizebox{0.38\columnwidth}{!}{\includegraphics{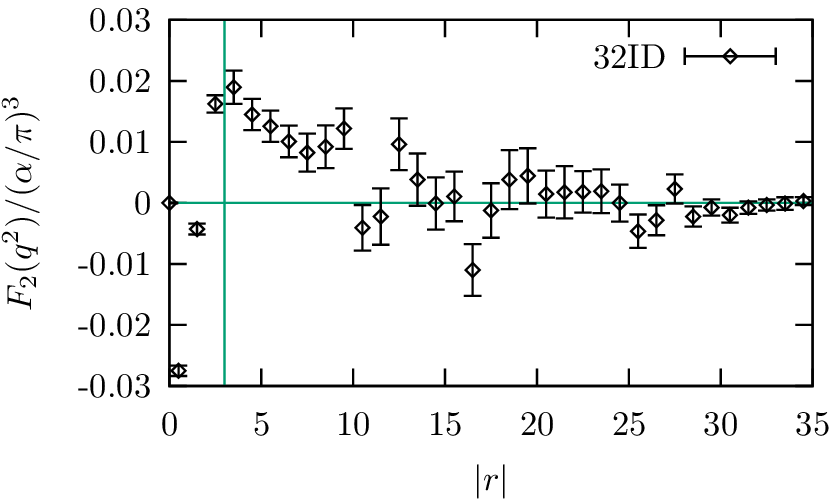}}\resizebox{0.38\columnwidth}{!}{\includegraphics{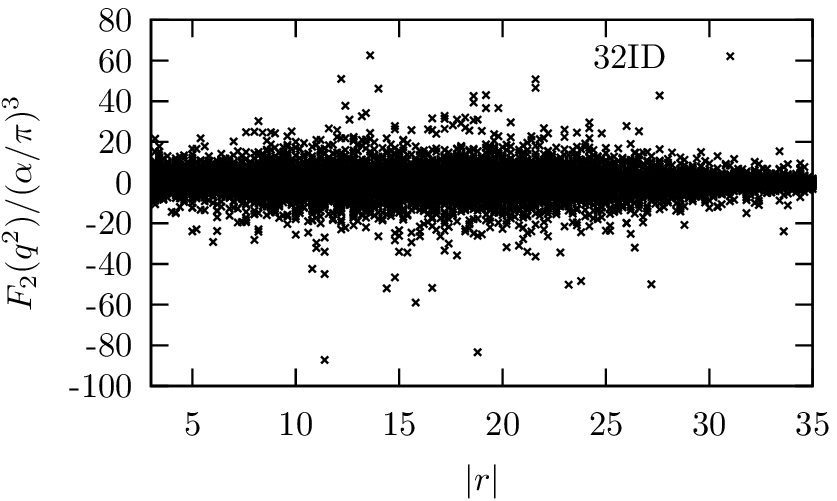}}
\\
    \resizebox{0.38\columnwidth}{!}{\includegraphics{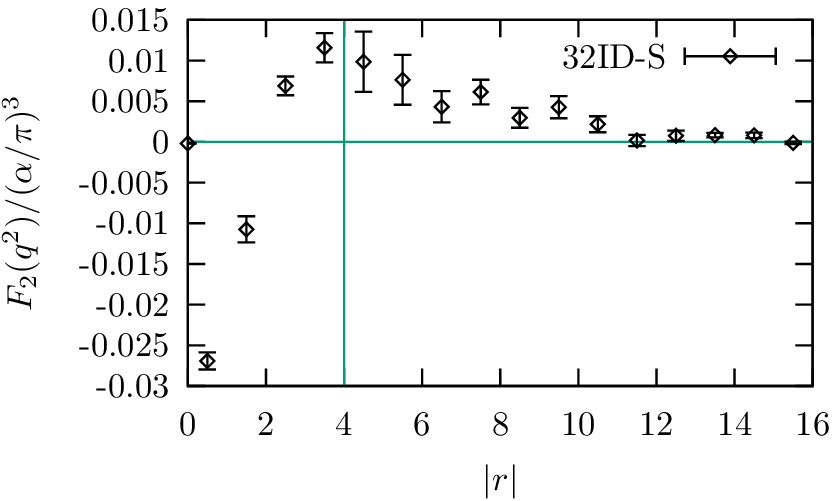}}\resizebox{0.38\columnwidth}{!}{\includegraphics{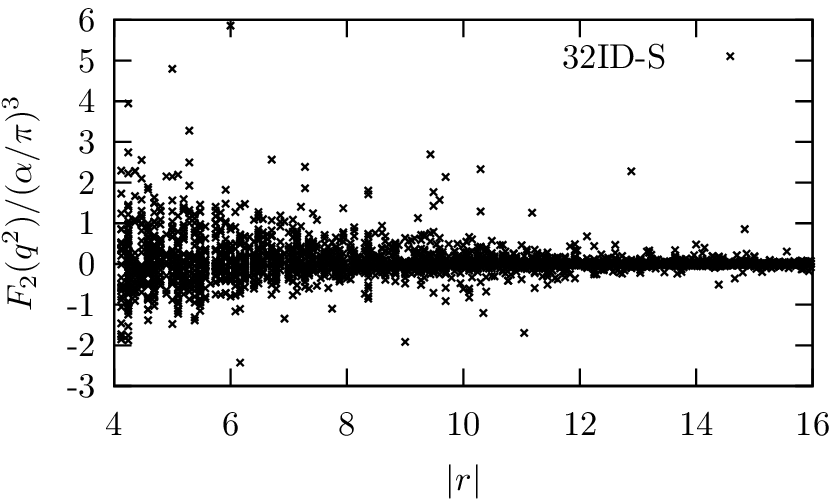}}
  \end{center}
\vskip -0.2 in
  \caption{The left column shows histograms of the contribution to $F_2$ from different separations $|r| = |x-y|$. The sum of all these points gives the final result for $F_2$.  The right column contains scatter plots of results for $F_2$ for all random point pairs, adjusted by their sampling weight. The average value of $F_2$ from all points gives the $|r|\ge r_{\max}$ portion of the final result.  The vertical lines in the left plots and the left-hand boundaries of the right plots indicate the value of $r_{\max}$.  The labels 16I, 24I, 24IL and 32ID indicate the ensembles given in Tab.~\ref{tab:confs}.}
  \label{fig:f2-ceil-r}
\end{figure}

We conclude the discussion of the exact propagator method at non-zero $q^2$ by examining two of the possible enhancements.  The first involves including two extra diagrams so that the external current is conserved on each configuration as was discussed in Sec.~\ref{sec:WI_per_config}.
The second can be viewed as an adaptation of the $M^2$ method discussed for the case of stochastic fields in Sec.~\ref{sec:M_2-method} to the exact propagator case.  In the present case we compute the needed sequential source quark propagators from $M$ locations of the point $x$ and then evaluate the contribution to $F_2$ from each of the $M(M-1)/2$ distinct pairs that can be formed from this set of $M$ points.

\subsubsection{Conserved current on each configuration}

We repeated the 32ID lattice computation with the same parameter choices but included all three diagrams in Fig.~\ref{fig:conserved_diags} in order to determine the value of this potential enhancement, described in subsection~\ref{sec:WI_per_config}.  The results are listed in Tab.~\ref{tab:32ID-cmp} as the ``Conserved'' method.  We find that although the cost per stochastic point is seven times larger than for the case that only one diagram is evaluated, this extra cost yields a marginal, over-all benefit in the reduction in noise.

\subsubsection{$M^2$ method}

We can analyze the effectiveness of this $M^2$ method for exact-photon propagators by comparing two different methods of estimating the statistical error that results from the long-distance contribution to $F_2$ coming from point-pairs with $r \ge r_{\mathrm{max}}$.  In Tab.~\ref{tab:32ID-cmp}, we list separately the results and errors from the short- and long-distance parts.  The errors are correct statistical errors computed from the variance of the average values obtained for each configuration.   However, we can also estimate a second long-distance error, denoted as ``ind-pair'' in the table, by assuming that the long-distance point-pairs are all completely independent even though on a given configuration they are simply different combinations of the same set of points.  If the correlations between these point pairs are significant, we should expect that the error obtained by treating them as independent and dividing the width of the distribution of results from these pairs by $\sqrt{N_{\mathrm{set}} N_{\mathrm{conf}} M(M-1)/2}$, will be less than the true error, determined by the first method described above.  From the table we can see that the error found by treating these $M(M-1)/2$ pairs as independent is only slightly smaller than the actual error which suggests a significant gain from evaluating the contribution of these $M(M-1)/2$ pairs.  Once again we see an $O(M^2)$ statistical advantage from the calculation of only $O(M)$ propagators.

\subsection{Moment method}
\label{sec:moment_numerical}

Here we present results that are obtained by using the best of the strategies discussed in Section~\ref{sec:strategy}.  Specifically we evaluate $F_2(q^2)$ at the point of interest $q^2=0$ using the moment method of Sec.~\ref{sec:moment}.  We also introduce the restriction $|z-x| \ge |x-y|$ and  $|z-y| \ge |x-y|$ explained at the end of Section~\ref{sec:WI_per_config} in order to more accurately sample the region where one of three vertices is far from the other two.  We use the 32ID ensemble lattice and increase the efficiency of the calculation by using the All-Mode-Averaging (AMA) method~\cite{Blum:2012uh, Shintani:2014vja} in which most of the propagator inversions are computed imprecisely and a small but more computationally expensive correction term is computed far less frequently.  We compute the short distance part up to $r_\text{max} = 5$ with the following samplings.  We compute point pairs with $|r| \le 1$ six times, $1 < |r| \le 2$ five times, $2 < |r| \le 3$ four times, $3 < |r| \le 4$ two times, $4 < |r| \le 5$ one time for each configuration. We use Eq.~\eqref{eq:moment_Xprod} in this computation and make use of its invariance under a larger set of discrete symmetries, including independent inversions of $x$, $y$, $z$, $t$, and the exchange of the $x$, $y$ and $z$ directions.

For the long distance part, we compute $512$ pairs per configuration.  In order to more precisely control the distributions of these long distance, $r > 5$, point pairs, we do not use the $M^2$ method in this calculation and instead chose the individual pairs so that their separation $r$ follows the probability distribution
\begin{eqnarray}
  P_\text{32ID} (r) & \propto & \frac{1}{|r|^4} e^{-0.05 |r|}.
\label{eq:moment_method_dist}
\end{eqnarray}
The approximate AMA results are computed using propagators that were obtained using only 100 conjugate gradient (CG) iterations.  We treat the AMA correction as a separate computation on the same set of configurations. For the short-distance part, we sum the contribution of the point pairs up to $r_\text{max} = 2$. We compute $48$ long-distance point pairs per configuration, using the same pair-separation distribution given in Eq.~\eqref{eq:moment_method_dist} for the long distance part of the AMA correction, but with $|r| > 2$.  On this restricted sample we compute the result from propagators computed using only 100 CG iterations and propagators computed with a residual of $10^{-8}$.

The results are presented in the final three rows of Tab.~\ref{tab:32ID-cmp}.  We use $m_\mu = 134$ MeV and a separation between the muon source and sink of $t_\text{sep} = 32$.  As in previous tables $\sqrt{\text{Var}} =\text{Err} \sqrt{N_{\mathrm{conf}} N_{\mathrm{prop}}}$ where the number of propagators computed per configuration, $N_{\mathrm{prop}}$, is defined as before.  In the moment method, for each point we compute $1$ point source propagator and $3$ sequential source propagators for each of the three spatial magnetic moment directions. Since the $\sqrt{\text{Var}}$ is based on the number of propagators computed, the reduction in $\sqrt{\text{Var}}$ seen between the ``Conserved'' and ``Mom. (aprox)'' rows of Tab.~\ref{tab:32ID-cmp} suggest that we get 40\% speed up from the moment method in addition to the gain in inversion speed that results from using the AMA approach.   Although we limit the approximate CG inversions to only $100$ iterations, compared with precise inversions which require $~1300$ iterations, the correction is very small. However, the variance of the correction is rather large, suggesting that the choice of $100$ approximate iterations may not be optimum.

In the results presented in Tab.~\ref{tab:32ID-cmp} we use local currents for the internal photons.  In the ``Exact'' and ``Conserved'' methods, we use the conserved current for the external photon, while in the moment method, we use a local current for the external photon.  The final row of Tab.~\ref{tab:32ID-cmp}, labeled ``Mom. (tot)'', gives the complete result from the moment method while the preceding two rows ``Mom. (approx)" and ``Mom. (corr)'' show separately the approximate AMA results and the needed correction term.  The ``SD'' and ``LD'' columns give the results from the pairs with $|r| \le r_{\mathrm{max}}$ and $|r| > r_{\mathrm{max}}$, respectively.  The ``ind-pair'' column gives the error that would be expected if the long-distance pairs were truly independent.  Note that the quantity $F_2(q^2)$ is computed at $q^2=(2\pi/L)^2$ for the first two rows and at $q^2=0$ for the final three rows.   The final error shown for the moment method on the fifth line of Tab.~\ref{tab:32ID-cmp} is obtained by applying the jackknife method to the sum of the approximate AMA result and the AMA correction term.  The resulting error is similar to what would be found were the statistical error on the approximate and correction terms computed separately and added in quadrature.

\begin{table}
  \begin{center}
    \begin{tabular}{lcccccccc}
      \hline
Method & $F_2/(\alpha / \pi)^3$ & $N_{\mathrm{conf}}$ & $N_{\mathrm{prop}}$ & $\sqrt{\text{Var}}$
& $r_\text{max}$ &SD     & LD     & ind-pair \\
      \hline
      \hline
      Exact			& $0.0693(218)$	& $47$ & $58 + 8 \times 16$	& $2.04$
                                & $3$      			& $-0.0152(17)$ 			& $0.0845(218)$	& $0.0186$\\
      Conserved		& $0.1022(137)$	& $13$ & $(58 + 8 \times 16) \times 7$    & $1.78$
				& $3$      			& $0.0637(34)$ 			& $0.0385(114)$	& $0.0093$\\
      Mom. (approx)	& $0.0994(29)$	& $23$ & $(217 + 512) \times 2 \times 4 $ & $1.08$
				& $5$			& $0.0791(18)$ 			& $0.0203(26)$	& $0.0028$\\
      Mom. (corr)	& $0.0060(43)$	& $23$ & $(10 + 48) \times 2 \times 4 $   & $0.44$
				& $2$			& $0.0024(6)$				& $0.0036(44)$	& $0.0045$\\
      Mom. (tot)		& $0.1054(54)$	& $23$ &                                  &       \\
      \hline
    \end{tabular}
  \end{center}
\caption{Results from three variants of the exact photon method obtained from the 32ID ensemble.  The first row, labeled ``Exact'', corresponds to the row labeled 32ID in Tab.~\ref{tab:psrc-results}.  The second row, labeled ``Conserved'' is similar except all three arrangements of the vertices $x$, $y$ and $z$ are combined insuring that the external current is conserved on each configuration.  The final three rows are obtained from the moment method and are explained in the text.}
\label{tab:32ID-cmp}
\end{table}

We should emphasize that the moment-method result given in the final line of Tab.~\ref{tab:32ID-cmp} is the most important numerical result presented in this paper.  It provides the cHLbL contribution (calculated directly at $q^2=0$) to $g-2$ for the muon with a 5\% statistical accuracy for the case of a pion with $m_\pi= 171$ MeV using a $(4.6~\mathrm{fm})^3$ spatial volume but with a relatively coarse lattice spacing $a$ with $1/a=1.378$ GeV.  More information about the conserved and moment method calculations presented in Tab.~\ref{tab:32ID-cmp} can be found in Fig.~\ref{fig:f2-ceil-r} where histograms and scatter plots are presented as functions of the separation of the two stochastically chosen points $x$ and $y$.

\begin{figure}
  \begin{center}
    \resizebox{0.48\columnwidth}{!}{\includegraphics{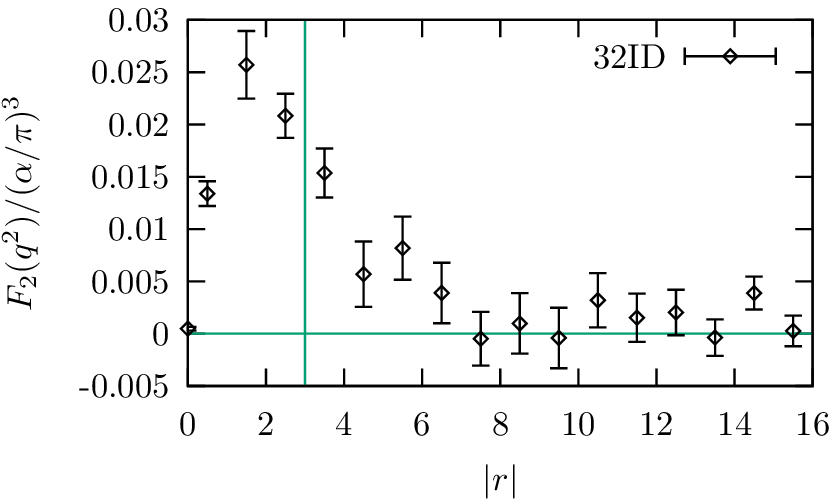}}\resizebox{0.48\columnwidth}{!}{\includegraphics{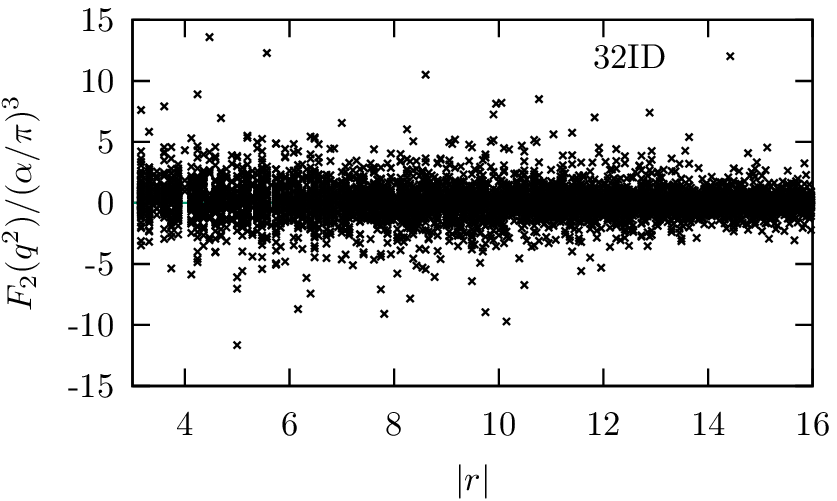}}
\\
    \resizebox{0.48\columnwidth}{!}{\includegraphics{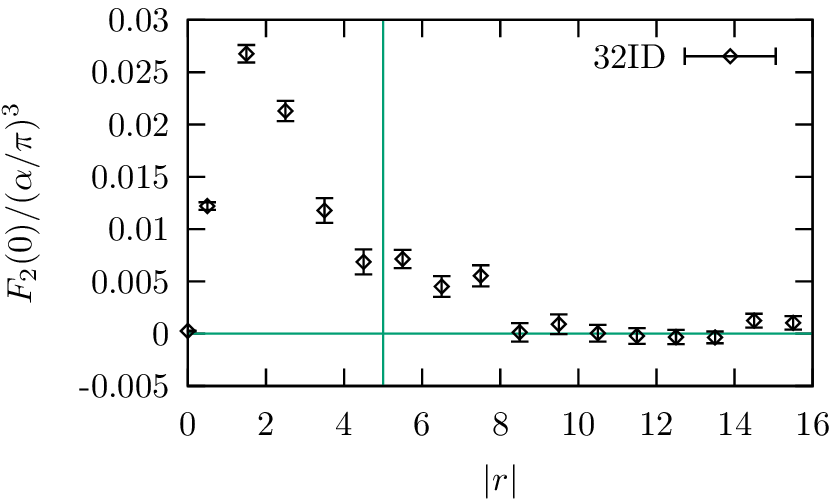}}\resizebox{0.48\columnwidth}{!}{\includegraphics{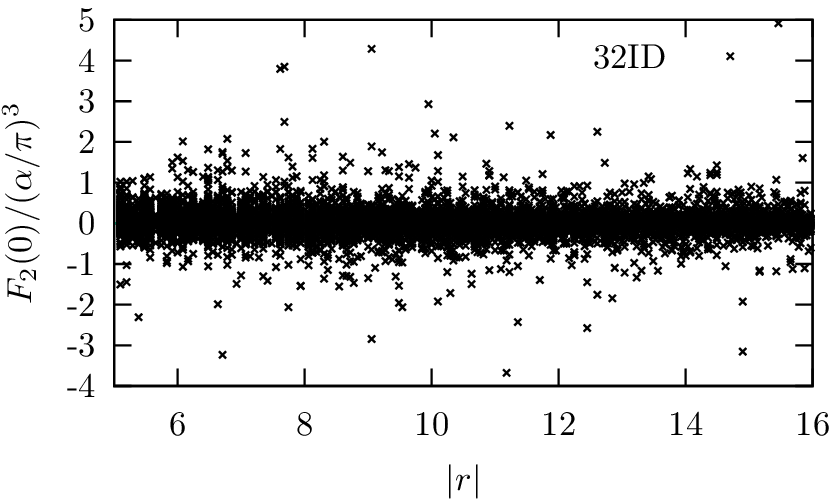}}
  \end{center}
  \caption{Histograms and scatter plots for the contribution to $F_2$ from different separations $|r| = |x-y|$ are shown in the left and right plots respectively, following the conventions used in similar, previous figures. The upper two plots are obtained using the conserved version of the exact photon method on the 32ID ensemble. The lower two plots are obtained using the moment method, but from approximate propagators each obtained from 100 CG iterations, again on the 32ID ensemble.}
  \label{fig:f2-ceil-r}
\end{figure}

As a final topic in this section we apply the conserved method and the moment method, with the restriction $|z-x| \ge |x-y|$ and $|z-y| \ge |x-y|$ that was described previously, to the 24I ensemble with $m_\mu a = 0.1$ in order to compare these methods with the original subtraction calculation~\cite{Blum:2014oka} which was carried out on the same ensemble with the same muon mass.  We compute the short distance part up to $r_\text{max} = 4$.  For $|r| \le 2$ we compute each independent direction two times while for $2 <|r| \le 4$ each independent direction is computed only once for each configuration. We take many discrete symmetries into account when summing over the short-distance part, including independent inversions of $x$, $y$, $z$, $t$, and exchanges of the $x$ and $y$ directions.  For the long-distance part, we did not use the $M^2$ method, but instead directly chose the probability distribution for the point pairs ($|r| > 4 $):
\begin{eqnarray}
  P_\text{24IL} (r) & \propto & \frac{1}{|r|^4} e^{-0.1 |r|} .
\end{eqnarray}

For the conserved method the propagators are computed with approximate inversions carried out to a precision of $10^{- 4}$.  (No correction term has been added.)  The number of propagators needed per configuration ($N_{\mathrm{prop}}$) is given by the sum of the number of point pairs times the twice the number of propagators computed per point.  For the conserved method, for each point we compute one point source propagator and six sequential source propagators, corresponding to the three external photon polarizations and two momentum directions.

\begin{figure}[t]
  \begin{center}
    \resizebox{0.9\columnwidth}{!}{\includegraphics{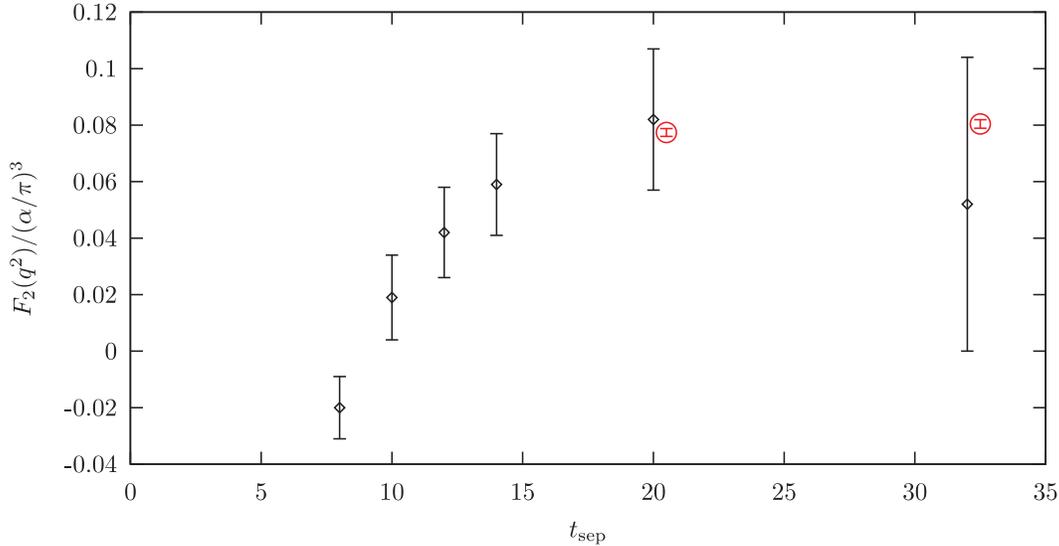}}
  \end{center}
\caption{A comparison of the results for $F_2(q^2)/(\alpha/\pi)^3$ obtained in the original lattice QCD cHLbL calculation~\cite{Blum:2014oka} (diamonds) with those obtained on the same gauge field ensemble using the moment method presented here (circles).  The points from the original subtraction method with $q^2=(2\pi/24)^2 = (457 \mathrm{MeV})^2$ were obtained from 100 configurations and the evaluation of 81,000 point-source quark propagators for each value of the source-sink separation $t_{\mathrm{sep}}$.  In contrast, the much more statistically precise results from the moment method required a combined 26,568 quark propagator inversions for both values of $t_{\mathrm{sep}}$ and correspond to $q^2=0$. The moment method value for $t_\text{sep} = 32$ is listed in Tab.~\ref{tab:24IL-conserved-moment}.}
  \label{fig:old_new_compare}
\end{figure}

For this implementation of the moment method we compute only the external momentum in $z$ direction, and external photon polarizations in $x$ and $y$ directions, so for each point we compute $1$ point source propagator and $2$ sequential source propagators for these two external photon polarizations.  This is slightly different (and less effective) than the approach used for the moment method given in Tab.~\ref{tab:32ID-cmp}.  The results are shown in Tab.~\ref{tab:24IL-conserved-moment} and a direct comparison between the $q^2=0$ results of the moment method (at two different muon source-sink separations) and the earlier $q^2=(2\pi/L)^2$ results of Ref.~\cite{Blum:2014oka} is shown in Fig.~\ref{fig:old_new_compare}.  As can be seen, a substantial improvement over the original calculation has been obtained.  In addition, the good agreement between the earlier results and the new results using the conserved current method, both at $q^2=(2\pi/L)^2$, provide a useful consistency check since these are two completely independent calculations.

\begin{table}[t]
  \begin{center}
    \begin{tabular}{lcccc}
      \hline
      Method & $F_2/(\alpha / \pi)^3$ & $N_{\mathrm{conf}}$ & $N_{\mathrm{prop}}$ & $\sqrt{\text{Var}}$\\
      \hline
      \hline
Conserved	& $0.0825(32)$ & $12$ & $(118 + 128) \times 2 \times 7$ & $0.65$\\
Mom.		& $0.0804(15)$ & $18$ & $(118 + 128) \times 2 \times 3$ & $0.24$ \\
      \hline
    \end{tabular}
  \end{center}
\caption{Results for $F_2(q^2)$ from applying the conserved and moment methods to the the 24IL ensemble with $m_\mu a = 0.1$ using a muon source-sink separation $t_\text{sep} = 32$.  As before,  $\sqrt{\text{Var}}=\text{Err} \sqrt{N_{\mathrm{conf}} N_{\mathrm{prop}}}$.   We use the  conserved current for the external photon and local currents for the internal photons for both methods.  The conserved results are for $q^2=(2\pi/L)^2$ while the moment methods gives a $q^2=0$ result.}
\label{tab:24IL-conserved-moment}
\end{table}

\subsection{QED light-by-light scattering results} \label{sec:QED_results}

In this section we present results for QED light-by-light scattering in which the quark loop discussed in the previous sections is replaced by a muon loop.  These calculations make use of the most effective of the numerical strategies discussed above: the use of exact photon propagators and the position-space moment method to determine $F_2$ evaluated at $q^2=0$.  Since these calculations are less computationally costly than those for QCD we can evaluate a number of volumes and lattice spacings (all specified with reference to the muon mass) and examine the continuum and infinite volume limits.   We can then compare our results, extrapolated to vanishing lattice spacing and infinite volume, with the known result calculated in standard QED perturbation theory~\cite{Laporta:1991zw, Jegerlehner:2009ry}.  This QED calculation both serves as a demonstration of the capability of lattice methods to determine such light-by-light scattering amplitudes and as a first look at the size of the finite-volume and non-zero-lattice-spacing errors.

\begin{figure}
  \begin{center}
    \resizebox{0.7\columnwidth}{!}{\includegraphics{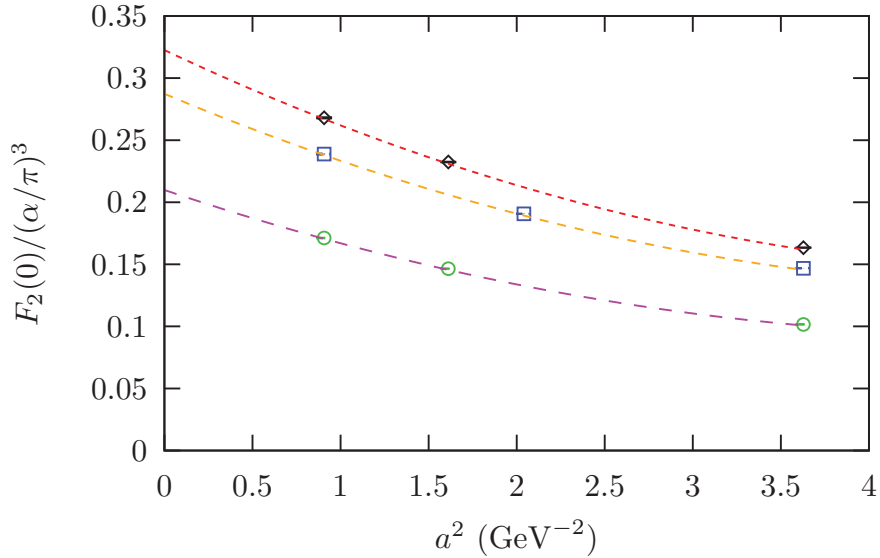}}
  \end{center}
\caption{Plots of our results for the connected light-by-light scattering contribution in QED to $F_2(0)$, known to be $0.371 \times \left(\alpha/\pi\right)^3$~\cite{Laporta:1991zw, Jegerlehner:2009ry}, as a function of $a^2$ expressed in GeV by assigning $m_\mu = 106$ MeV.  This is done for three choices of the physical lattice size $L= 11.9\mathrm{fm}$ (diamonds), $8.9 \mathrm{fm}$ (squares) and $5.9 \mathrm{fm}$ (circles).  The curves shown are quadratic functions of $a^2$ chosen to pass through the three points for each physical volume.  The coefficients for each of these fits are listed in Tab.~\ref{tab:QED_fit_lin}.}
\label{fig:QED_a2_extrapolation}
\end{figure}

In Fig.~\ref{fig:QED_a2_extrapolation} we show results for $F_2(0)$ computed for three different lattice spacings, {\it i.e.} three different values of the input muon mass in lattice units, but keeping the linear size of the system fixed in units of the muon mass.  The data shown in Fig.~\ref{fig:QED_a2_extrapolation} are also presented in Tab.~\ref{tab:QED_results}.   We use two extrapolation methods to obtain the continuum limit.  The first, shown in the figure, uses a quadratic function of $a^2$ to extrapolate to $a^2 = 0$.  The second makes a linear extrapolation to $a^2 = 0$ using only the two left-most points for each of the three values of $m_\mu L$.  The coefficients for the quadratic-in-$a^2$ fits shown in Fig.~\ref{fig:QED_a2_extrapolation} as well as those for the linear-in-$a^2$ fits are given in tabular form in Tabs.~\ref{tab:QED_fit_quad} and \ref{tab:QED_fit_lin}.

\begin{table}[t]
  \begin{center}
    \begin{tabular}{llccrrr}
      \hline
      Vol	&\multicolumn{1}{c}{$m_\mu$} & $r_{\text{max}}$ & $p (x)$ &
      $N_{\mathrm{pair}}$ & $F_2(0) / (\alpha / \pi)^3$ & SD\\
      \hline\hline
      $16^3\times64$	& 0.2 	& 5 & $\frac{\exp (- 0.4 | x |)}{| x |^4}$ & 1024 & $0.1016(1)$ & $0.1000$\\
      $24^3\times96$ 	& 0.1333	& 6 & $\frac{\exp (- 0.25 | x |)}{| x |^4}$ & 86 & $0.1465(3)$ & $0.1428$\\
      $32^3\times128$	& 0.1 & 6 & $\frac{\exp (- 0.2 | x |)}{| x |^4}$ & 194 & $0.1712(3)$ & $0.1624$\\
      \hline
      $24^3\times96$	& 0.2 & 6 & $\frac{\exp (- 0.4 | x |)}{| x |^4}$ & 80 & $0.1468(1)$ & $0.1451$\\
      $32^3\times128$	& 0.15 & 6 & $\frac{\exp (- 0.3 | x |)}{| x |^4}$ & 50 & $0.1907(2)$ & $0.1863$\\
      $48^3\times192$	& 0.1 & 6 & $\frac{\exp (- 0.2 | x |)}{| x |^4}$ & 152 & $0.2388(5)$ & $0.2243$\\
      \hline
      $32^3\times128$	& 0.2 & 5 & $\frac{\exp (- 0.4 | x |)}{| x |^4}$ & 276 & $0.1634(2)$ & $0.1613$\\
      $48^3\times192$	& 0.1333 & 6 & $\frac{\exp (- 0.25 | x |)}{| x |^4}$ & 189 & $0.2324(3)$ & $0.2291$\\
      $64^3\times128$	& 0.1 & 6 & $\frac{\exp (- 0.2 | x |)}{| x |^4}$ & 184 & $0.2680(5)$ & $0.2592$\\
      \hline
    \end{tabular}
  \end{center}
  \caption{A list of the input parameters, weights and numerical results for our QED calculations using the moment method.  The right-most column shows the very accurate results from the short-distance, $|r|\le r_{\max}$ region.  These results are plotted in Fig.~\ref{fig:QED_a2_extrapolation}.}
\label{tab:QED_results}
\end{table}

\begin{table}
  \begin{center}
    \begin{tabular}{cc}
      \hline
      $L/\mathrm{fm}$ & $F_2(0) / (\alpha / \pi)^3$ \\
      \hline\hline
      5.9  & $0.2099(12) - 0.0478(13)(a~\mathrm{GeV})^2 + 0.0049(3)(a~\mathrm{GeV})^4$ \\
      8.9  & $0.2873(13) - 0.0595(11)(a~\mathrm{GeV})^2 + 0.0056(2)(a~\mathrm{GeV})^4$ \\
      11.9 & $0.3226(17) - 0.0669(17)(a~\mathrm{GeV})^2 + 0.0062(4)(a~\mathrm{GeV})^4$ \\
      \hline
    \end{tabular}
  \end{center}
\caption{Functions quadratic in $a^2$ which fit the data shown in Fig.~\ref{fig:QED_a2_extrapolation}. The results from these fits at $a^2=0$ are plotted in Fig.~\ref{fig:QED_V_extrapolation}.}
  \label{tab:QED_fit_quad}
\end{table}

\begin{table}
  \begin{center}
    \begin{tabular}{cc}
      \hline
      $L/\mathrm{fm}$ & $F_2(0) / (\alpha / \pi)^3$ \\
      \hline\hline
      5.9  & $0.2030(8) - 0.0357(6)(a~\mathrm{GeV})^2 $ \\
      8.9  & $0.2773(9) - 0.0432(5)(a~\mathrm{GeV})^2 $ \\
      11.9 & $0.3138(12) - 0.0515(9)(a~\mathrm{GeV})^2 $ \\
      \hline
    \end{tabular}
  \end{center}
\caption{Functions linear in $a^2$ which can be used to extrapolate the data shown in Fig.~\ref{fig:QED_a2_extrapolation} to $a^2=0$.  The results from these fits at $a^2=0$ are plotted in Fig.~\ref{fig:QED_V_extrapolation}.}
  \label{tab:QED_fit_lin}
\end{table}

\begin{figure}
  \begin{center}
    \resizebox{0.7\columnwidth}{!}{\includegraphics{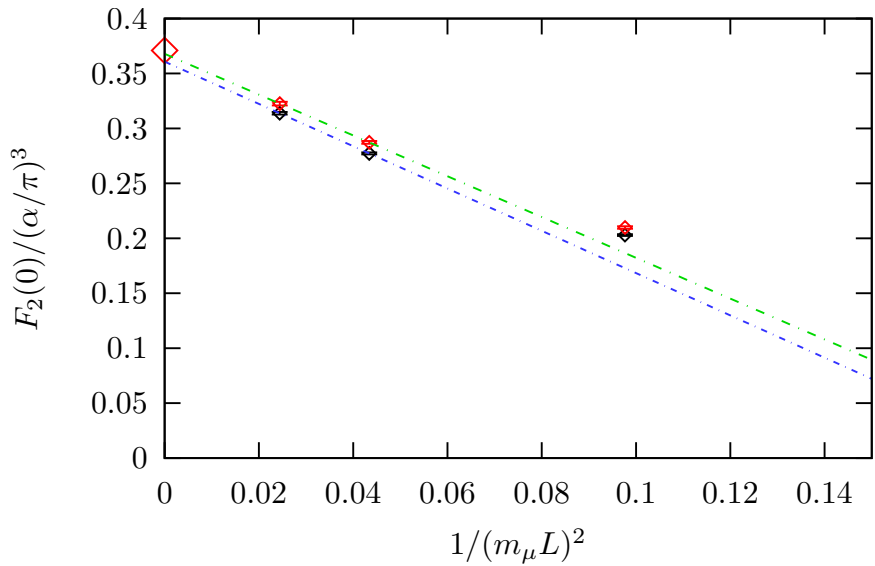}}
  \end{center}
  \caption{Results for $F_2(0)$ from QED connected light-by-light scattering.  These results have been extrapolated to the $a^2\to 0$ limit using two methods.  The upper points use the quadratic fit to all three lattice spacings shown in Fig.~\ref{fig:QED_a2_extrapolation} while the lower point uses a linear fit to the two left most points in that figure.  Here we extrapolate to infinite volume using the linear fits shown to the two, left-most of the three points in each case. }
  \label{fig:QED_V_extrapolation}
\end{figure}

In Fig.~\ref{fig:QED_V_extrapolation} we plot the $a^2=0$ values that result from the quadratic fit to the $a^2$  dependence given in Tab.~\ref{tab:QED_fit_quad} as a function of $1/(m_\mu L)^2$ along with the original perburbative result for these QED terms.  There is clearly good agreement between an extrapolation linear in $1/(m_\mu L)^2$ using the two left-most points and the known perturbative result.  These fitting results, shown as functions of $m_\mu L$ are summarized in the following equations:
\begin{eqnarray}
\left[F_2(0)\right]_{\mathrm{quad}}/(\alpha/\pi)^3 &=& 0.3679(42) - 1.86(11) / (m_\mu L)^2,
\label{eq:QED_quad} \\
\left[F_2(0)\right]_{\mathrm{lin}}/(\alpha/\pi)^3 &=& 0.3608(30) - 1.92(8)/ (m_\mu L)^2,
\label{eq:QED_lin} \\
\left[F_2(0)\right]_{\mathrm{PT}}/(\alpha/\pi)^3 &=& 0.3710052921 \label{eq:QED_PT},
\end{eqnarray}
where the errors shown in Eqs.~\eqref{eq:QED_quad} and \eqref{eq:QED_lin} are statistical only and the perturbative result is given in Eq.~\eqref{eq:QED_PT}.  We find very satisfactory agreement between the results from standard perturbation theory and the lattice results extrapolated to the continuum and infinite volume limits.

\section{Conclusion and Outlook}
\label{sec:conclusion}

In this paper we have extended the lattice field theory methods introduced in Ref.~\cite{Blum:2014oka}, increasing the computational efficiency by more than two orders of magnitude and allowing the calculation of the $q^2$-dependent form factor $F_2(q^2)$ directly at $q^2=0$ instead of at $(2\pi/L)^2$, the smallest, non-zero momentum accessible in finite volume.  To demonstrate the correctness of our methods we have studied the light-by-light scattering contribution within QED, arising when the internal loop is a muon, working at three values for the lattice spacing and three volumes.  By extrapolating to vanishing lattice spacing and infinite volume we obtain a result which agrees with the analytic result within 2\%, an accuracy expected from a combination of statistical and extrapolation uncertainties.

The most successful approach uses exact, analytic formulae for the three photon propagators that appear in the HLbL amplitude and the standard methods of lattice QCD.  In contrast with normal perturbative methods, much of the calculation is performed in position space and stochastic methods are only introduced to sample position-space sums, reducing the computational cost so that it grows proportional to the space-time volume instead of its cube.  Because of the structure of the amplitude being computed, we can identify a specific space-time position within the hadronic part of the amplitude and use that location as the origin to obtain the anomalous magnetic moment from what is essentially a classical spatial moment of the quantum distribution of current.

These new methods are used to obtain a result for the cHLbL contribution to $g_\mu-2$ from a relatively coarse, $32^3\times64$ ensemble with $1/a=1.38$ GeV, spatial extent $L=4.6$ fm and pion mass $m_\pi=171$ MeV:
\begin{equation}
  \frac{(g_\mu-2)_{\mathrm{cHLbL}}}{2} = (0.1054 \pm 0.0054) (\alpha/\pi)^3 = (132.1 \pm 6.8)\times 10^{-11}.
\label{eq:result}
\end{equation}
which can be compared to the conventional model-dependent result for the complete HLbL contribution to $g_\mu-2$ of $(105\pm 26)\times 10^{-11}$ and the difference between the current experimental result and the standard model prediction (excluding the HLbL component) of $(354\pm86)\times 10^{-11}$.   Equation~\eqref{eq:result} shows only the statistical error.  There are significant systematic errors associated with the unphysical pion mass, the non-zero lattice spacing and the finite volume that have been used in this calculation.  These systematic errors are at present insufficiently well understood to be reliably estimated.  A particularly important systematic errors comes from the omission of the quark-disconnected contributions, which play an important role in the phenomenological estimates.  Thus, the comparison of the result in Eq.~\eqref{eq:result} with experiment serves only to give a context for the size of the present statistical errors.

In Section~\ref{sec:studies} we have presented a series of numerical tests of many of the different methods that were explored while developing the methods that were finally used to obtain the result in Eq.~\eqref{eq:result}.  We hope that some of these may be useful in the future for the efficient calculation of other quantities that involve a combination of QED and QCD, a relatively new area where there are many new directions to explore.

While the results presented here required modest computational resources, the result for the cHLbL contribution to $g_\mu-2$ requires substantially increased statistics as the pion mass decreases to its physical value.   However, based on the performance of the methods described here we expect that calculations at physical pion mass are practical on current leadership class computers and a calculation on an (5.5~fm$)^3$ volume with $1/a=1.73$ GeV is currently underway.  This calculation requires a $48^3\times 96$ lattice volume and a follow-on calculation with a smaller lattice spacing and a corresponding $64^3\times 128$ volume may also be possible, allowing a continuum limit to be evaluated.  Controlling the effects of finite volume and including the contributions of disconnected diagrams are more difficult but are being actively pursued.

\section{Acknowledgments}

We would like to thank our RBC and UKQCD collaborators for helpful discussions and critical software and hardware support.  The calculations reported here were carried out on the BG/Q machines at the RIKEN BNL Research Center and the Brookhaven National Laboratory.  T.B.~is supported by U.S. DOE grant \#DE-FG02-92ER41989. N.H.C.~and L.J.~are supported in part by U.S. DOE grant \#de-sc0011941.  M.H.~is supported by Grants-in-Aid for Scientific Research \#25610053. T.I.~and C.L.~are supported in part
by US DOE Contract \#AC-02-98CH10886(BNL). T.I.~is also supported by Grants-in-Aid for Scientific Research \#26400261. 

\appendix

\section{Avoiding lattice artifacts in the HLbL amplitude}
\label{sec:conserved-vs-local}

In standard continuum perturbation theory the Feynman graphs which enter the HLbL contribution to $g_\mu-2$ contain no divergences beyond the usual mass, wave function and coupling constant renormalizations that result from either the QED or QCD interactions.  In fact, because of the limited topologies for the photon couplings which appear in these HLbL amplitudes, even these standard QED renormalizations are not required.   However, when a lattice regulator is used, the choice of electromagnetic couplings may change this situation.  Wilson's formulation of lattice gauge theory introduces couplings between the quarks and gluons which explicitly preserve the Yang-Mills gauge symmetry even at finite lattice spacing and guarantees that gauge-non-invariant counter terms will not be needed to ensure that the lattice theory has a continuum limit.

Following the same strategy, we can avoid the appearance of new, unwanted short-distance contributions in a HLbL lattice calculation by introducing quark-photon and muon-photon couplings which are invariant under QED gauge symmetry.  This is quite manageable if a single photon is to be coupled to a muon or quark line: we can introduce the conserved lattice current which contains fermion fields evaluated at both ends of the given lattice link associated with the current operator.  However, if two or three photons are coupled to the same fermion line then the non-locality of the conserved current used to couple the first photon requires that additional two- and three-photon vertices be introduced if electromagnetic gauge invariance is to be preserved.  The resulting calculation can still be performed but at the cost of considerable complexity.

In this appendix we will demonstrate that new $O(1)$ lattice artifacts can be avoided in the case of the HLbL amplitude by the simple precaution of using the conserved lattice current when coupling the external photon to the quark loop.  The other six electromagnetic couplings can be given by the standard local current, provided the six necessary $Z_V$ renormalization factors are introduced.   The use of the conserved current for the external photon is only needed for the connected graph.  For the disconnected HLbL amplitudes the simpler local current can be used for all photon couplings.

\begin{figure}
\centering
\includegraphics[width=0.45\linewidth]{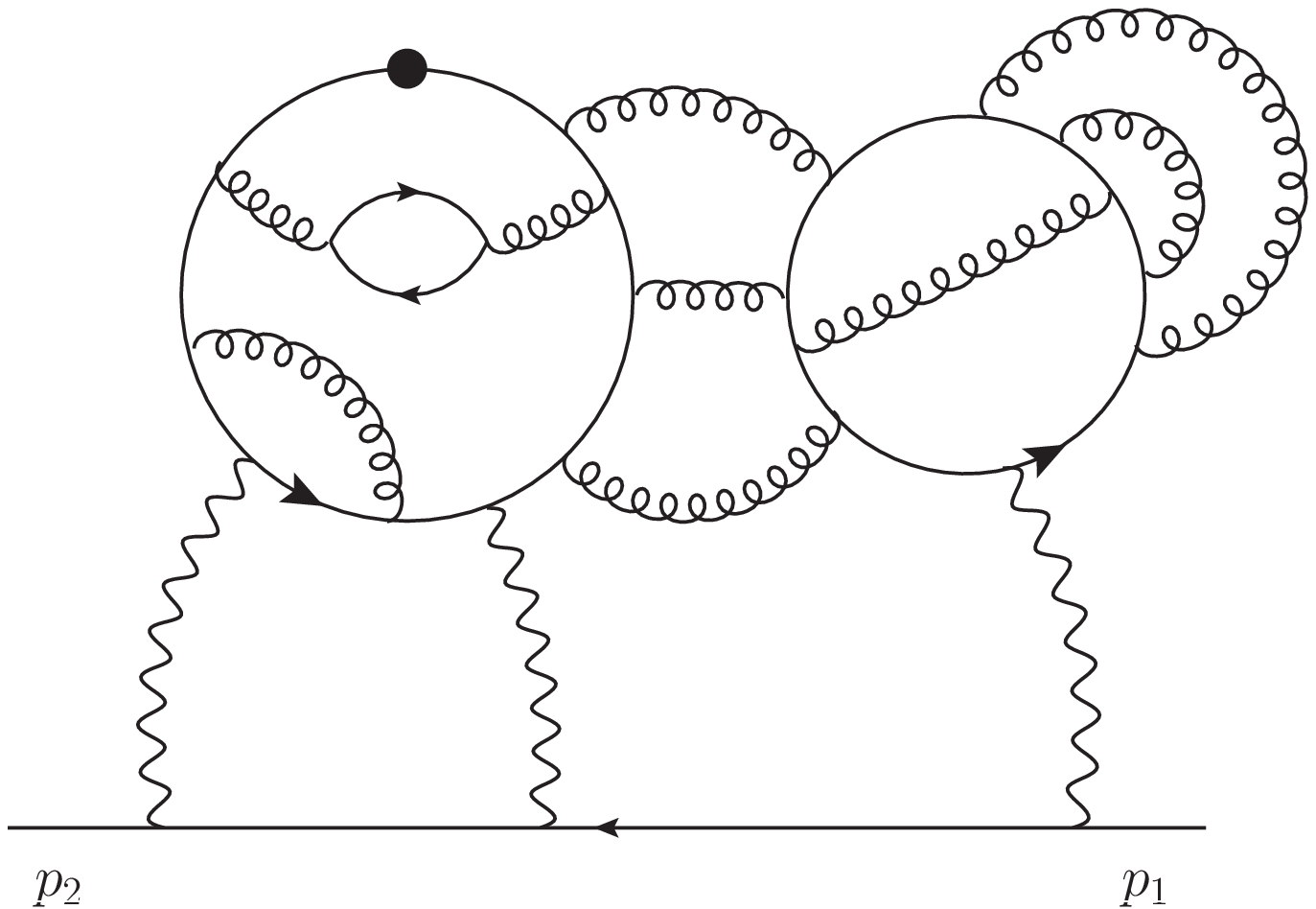}
\includegraphics[width=0.45\linewidth]{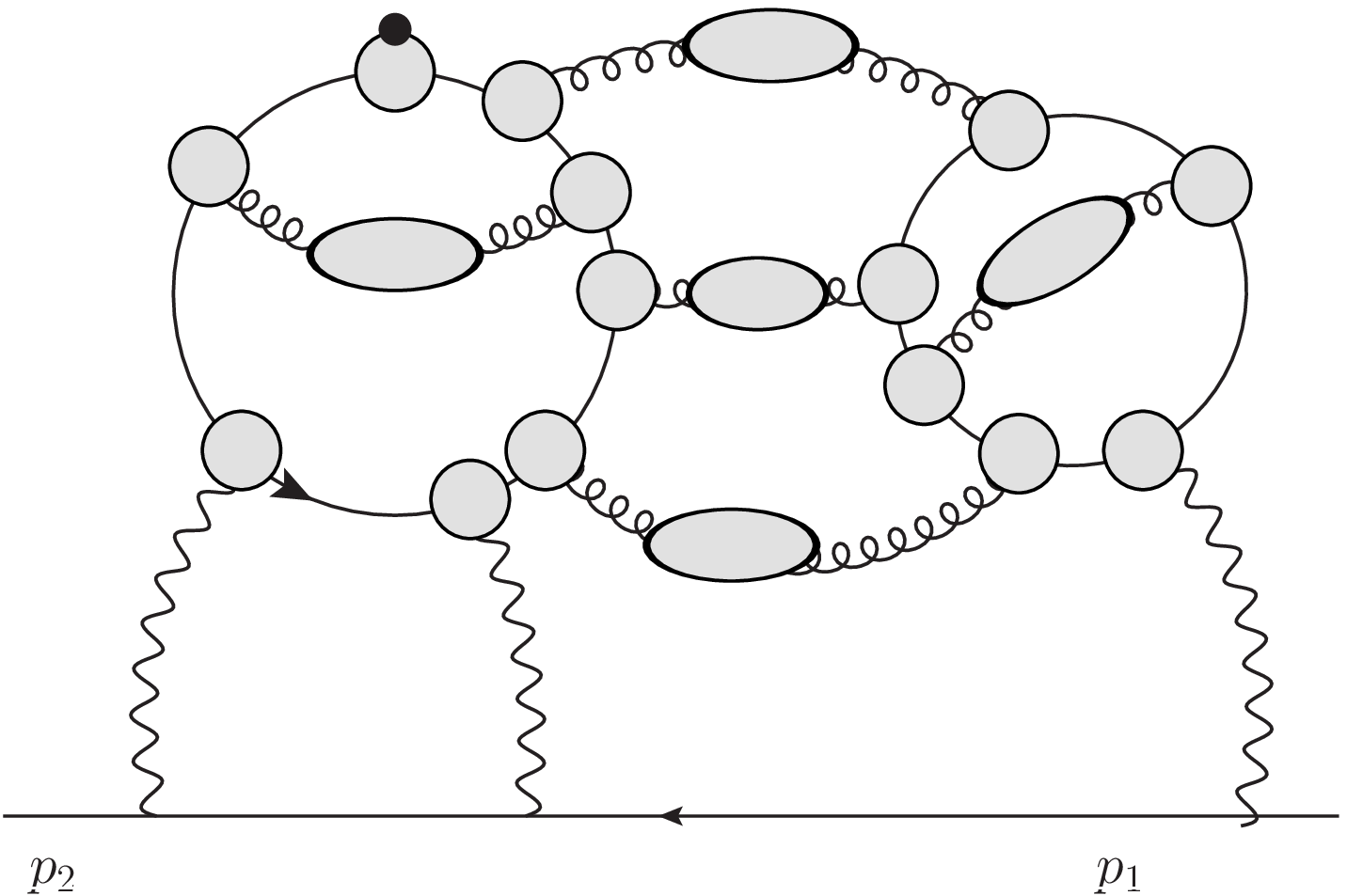}
\caption{The left-hand graph shows a sample QCD+QED diagram contributing to the HLbL amplitude.  The black dot in this diagram represents the current to which the external photon couples.  The right-hand graph shows the skeleton graph to which this sample graph contributes.  Here the shaded disk with the black dot on its circumference represents the full vertex function containing the current to which the external photon couples.}
\label{fig:skeleton}
\end{figure}

The absence of new short distance contributions when a local current is used for all internal photon couplings in a lattice-regulated calculation of HLbL can be seen by examining the HLbL amplitude in a Feynman perturbation theory expansion carried out to arbitrary order in the QCD coupling.  A convenient approach organizes the QCD perturbation theory into skeleton graphs and analyzes each skeleton graph~\cite{Bjorken:1965zz, ZinnJustin:2002ru}.  Recall that a skeleton graph in this context will be a graph with three internal photon lines and arbitrary quark and gluon lines subject to the restriction that no self-energy or proper vertex subgraphs appear.  Each vertex in such a skeleton graph represents a sum over all one-particle irreducible QCD vertex graphs.  Likewise each propagator in such a skeleton graph represents a sum over all QCD gluon or quark self-energy diagrams.   In Fig.~\ref{fig:skeleton} we show a sample HLbL graph and the corresponding skeleton graph.

Each such skeleton graph can be expanded into a sum of ordinary graphs by replacing each vertex and propagator by the corresponding sums over all vertex and propagator subgraphs.  Likewise a general graph can be identified with a skeleton graph if each vertex and self-energy subgraph appearing in that general graph is replaced by a simple vertex or propagator.  It can be shown that this process yields a unique skeleton graph independent of the order in which this replacement is made provided that the entire graph is not itself a self-energy graph, which it is not in the present HLbL case.  In a standard skeleton graph expansion the three internal photon propagators may themselves be part of a proper vertex or self-energy subgraph and would then not appear in the final skeleton graph.  However, for the HLbL case where each internal photon line is coupled to the single muon line which passes through the diagram, the only vertex or self-energy subgraph which contains one or more internal photon lines is the entire graph.

We will now show that each of the six internal photon vertices in this skeleton expansion can be accurately implemented if the internal photon is coupled to $Z_V$ multiplied by the local lattice current for each of the vertex subgraphs represented by that vertex in the skeleton graph expansion.  This will be the case if the momentum carried by each of the three external lines connected to that vertex is small compared to the regulator scale, which in our case is the inverse lattice spacing $1/a$.  It is for such physical-scale momenta that $Z_V$ times the local current and the conserved lattice current will agree.  Thus, we need to show that all skeleton subgraphs which contain such a complete internal photon vertex have a negative degree of divergence.   Such a convergent character for all graphs in which each internal photon vertex appears will guarantee that when the momenta entering that vertex is of order $1/a$, this will correspond to a momentum integration  region which is suppressed by at least two inverse powers of $1/a$ which results in a small, $O(a^2)$ error.

There are two types of skeleton graphs with a non-negative degree of divergence.  The first is the entire HLbL graph itself which as a vertex graph has zero degree of divergence.  However, for the case of the magnetic form factor $F_2$ being examined here, we are considering a term which is even under conjugation with $\gamma^5$.  Such a chirality changing amplitude will vanish unless an explicit factor of the muon mass is present and the presence of such a mass factor implies that the graph has degree of divergence $-1$ or smaller, guaranteeing suppression of the momentum region when all internal lines carry large momenta.

The other type of subgraph, which is neither a vertex nor a self-energy subgraph, but which has a potentially non-negative degree of divergence contains an internal quark loop coupled to four gluon or photon lines which are external lines of that subgraph.   In a gauge-invariant regularization scheme in which these gluons and photons couple to conserved currents, the corresponding Ward identities will guarantee that each of these currents is transverse which requires that the entire amplitude contains two or more explicit factors of the momenta carried by these four external gluons or photons.  The presence of these momentum factors reduces the zero degree of divergence of such a graph with four external boson lines, resulting in a negative degree of divergence.  Since each gluon couples to a conserved current which guarantees convergence of the subgraph, the only subgraphs at issue are those with four external photon lines.

Such subgraphs do appear in the HLbL amplitude and correspond to a quark loop with general internal gluon couplings but the only external vertices possessed by that subgraph are those of the three internal photons and the external current.  Thus, each such subgraph will have zero degree of divergence unless we require that one of these four couplings involve an exactly conserved current.  Thus, our choice that the photon external to the entire HLbL graph couple to a conserved current guarantees that this is the case.  Under these circumstances an explicit external momentum factor must be present and the subgraph must have a negative degree of divergence.  Note that this class of diagram which can be made convergent by the introduction of the conserved external current corresponds only to the connected, cHLbL case studied in this paper.  Such a conserved current coupling is not required for any of the disconnected graphs.

In this discussion we have assumed that the three internal photons couple to the quark and muon lines through a local current.  We have not been concerned about the short distance form of this local current since this will only affect the form of the coupling when large momentum flows through the vertex given by that current.  For a non-conserved local current this will act only to change the normalization of the current, an effect which is corrected by the introduction of the factor of $Z_V$.   We can also include more complex couplings for the internal photons without changing the final result.  For example if additional dimension-six, two-quark, three-photon couplings are introduced, the degree of divergence of these subgraphs will be increased and could become non-negative.  Such a dimension-six vertex would result in subgraphs with two quark, three photon external lines with degree of divergence increased from $-2$ to 0.  However, the factor of $a^2$ that must accompany such a dimension-six lattice operator would ensure that its effects would vanish as $a^2$ in the continuum limit.

\section{Conventional interpretation of moment formula}
\label{sec:classical_connection}

Equation~\eqref{eq:moment_Xprod} derived in Sec.~\ref{sec:moment} provides a very effective way to obtain $g_\mu-2$ from a first moment of the finite-volume cHLbL amplitude evaluated directly at zero-momentum transfer.  In this appendix we provide additional context for this equation by showing its relation to the conventional formula given in Eq.~\eqref{eq:classical} for the magnetic moment resulting from a localized static current distribution.  We begin by repeating Eq.~\eqref{eq:moment_Xprod}:
\begin{eqnarray}
\frac{F_2(0)}{2 m_\mu}\overline{u}(\vec 0,s')\,\vec\Sigma\,  u(\vec 0,s)  &=&
\frac{1}{2}\sum_{r,z,x_\op} \vec{x}_\op \times i\overline{u}(\vec 0,s')\, \vec{\mathcal{F}}^C \left(\frac{r}{2},-\frac{r}{2},z,x_\op\right) u(\vec 0,s).
\label{eq:moment_Xprod-copy}
\end{eqnarray}
While this equation is suggestive of the conventional Eq.~\eqref{eq:classical} for the magnetic moment there are three significant differences: i)  An internal coordinate in the Feynman amplitude on the right-hand side of Eq.~\eqref{eq:moment_Xprod-copy}, the variable $w=(x+y)/2$, is fixed at zero when it should be integrated over space-time in a perturbative evaluation of the matrix element of the current $\vec J(x_\op)$ in Eq.~\eqref{eq:classical}, ii) The time coordinate of the current, $(x_\op)_0$ is integrated instead of being held fixed, and iii) The factor of $1/V$ which is required if the initial and final muon states are to be properly normalized is missing.  As we will see these three differences between Eqs.~\eqref{eq:moment_Xprod} or \eqref{eq:moment_Xprod-copy} and Eq.~\eqref{eq:classical}, mutually compensate.

The first step in this demonstration exploits the symmetry of $\vec{\mathcal{F}}^C \left(\frac{r}{2},-\frac{r}{2},z,x_\op\right)$ under time translation to subtract $(x_\op)_0$ from each of the four time arguments in Eq.~\eqref{eq:moment_Xprod-copy}. This step will result in the external current being evaluated at $t=0$, an easily absorbed shift in the summation variable $z_0$ and the appearance of two independent summations over the time arguments of the points $x$ and $y$, allowing us to write Eq.~\eqref{eq:moment_Xprod-copy} as
\begin{eqnarray}
\frac{F_2(0)}{2 m_\mu}\overline{u}(\vec 0,s')\,\vec\Sigma\,  u(\vec 0,s)  && \label{eq:moment_Xprod-2} \\
&& \hskip -0.5 in
=\frac{1}{2}\sum_{\vec r, x_0, y_0 \atop z, \vec{x}_\op} \vec{x}_\op \times i\overline{u}(\vec 0,s')\, \vec{\mathcal{F}}^C \left((x_0,\frac{\vec r}{2}),(y_0, -\frac{\vec r}{2}),z,(0,\vec{x}_\op)\right) u(\vec 0,s), \nonumber
\end{eqnarray}
where we have written the previous $x$ and $y$ vertices as the four-vectors $(x_0,\frac{\vec r}{2})$ and $(y_0, -\frac{\vec r}{2})$ respectively and absorbed the $(x_\op)_0$ shift into the summation variable $z$.

Next we turn to the conventional formula, adapted to our quantum mechanical circumstances:
\begin{eqnarray}
\langle\psi'|\psi\rangle (\vec \mu)_{s',s} = \frac{1}{2} \sum_{\vec{x}_\op} \vec{x}_\op \times \left\{\sum_{\vec p\,' \vec p} \widetilde{\psi}'(\vec p\,')^* \Bigl\langle\mu(\vec p\,',s')\Bigl|\vec J(0, \vec x_\op)\Bigr|\mu(\vec p,s)\Bigr\rangle \widetilde{\psi}(\vec p)\right\},
\label{eq:moment_Xprod-3}
\end{eqnarray}
where $\widetilde{\psi}'(\vec p\,')$ and  $\widetilde{\psi}(\vec p)$ are momentum-space wave functions that describe initial and final muon states which are localized at the origin, which itself is chosen to be far from the walls of the large, finite volume in which the calculation is being performed.  (The wave functions $\widetilde{\psi}'(\vec p\,')$ and  $\widetilde{\psi}(\vec p)$ are normalized to $1/V$ to compensate for the states $|\mu(\vec p',s')\rangle$ and $|\mu(\vec p,s)\rangle$ being unnormalized plane waves.)  A  non-relativistic form has been assumed for the expression on left-hand side,  Finally we can recover Eq.~\eqref{eq:moment_Xprod-2} from Eq.~\eqref{eq:moment_Xprod-3} if we replace the matrix element between momentum eigenstates by the Feynman amplitude $i\overline{u}(\vec p\,',s')\vec{\mathcal{F}}^C(x,y,z,x_\op)u(\vec p , s) $ that appears in Eq.~\eqref{eq:moment_Xprod-2}:
\begin{equation}
\langle\psi'|\psi\rangle (\vec \mu)_{s',s}
 = -\frac{e}{2} \sum_{\vec{x}_\op}  \vec{x}_\op \times \left\{\sum_{\vec p\,' \vec p} \widetilde{\psi}'(\vec p\,')^* \sum_{\vec w} i\overline{u}(\vec p\,',s') \vec{\mathcal{F}}^C \Bigl(\vec w, (0, \vec x_\op)\Bigr)u(\vec p , s) \widetilde{\psi}(\vec p)\right\},
\end{equation}
where for clarity we display only the internal vector $\vec w = (\vec x +\vec y)/2$ in addition to $x_\op$.  We can use the translational covariance of $\mathcal{F}$ to extract the variable $\vec w$, rename the shifted variable $\vec x_\op-\vec w$ to simply $\vec{x}_\op$ and invoke current conservation to drop the added $\vec w$ that will appear in the left-hand factor of $\vec x_\op$ when this renaming is done.  We obtain:
\begin{eqnarray}
\langle\psi'|\psi\rangle (\vec \mu)_{s',s} &&\\
&&\hskip -0.5 in = -\frac{e}{2} \sum_{\vec{x}_\op}  \vec{x}_\op \times \left\{\sum_{\vec p\,' \vec p} \widetilde{\psi}'(\vec p\,')^* \sum_{\vec w} e^{i\vec w \cdot(\vec p - \vec p\,')} i\overline{u}(\vec p\,',s') \vec{\mathcal{F}}^C \Bigl(\vec 0, (0, \vec x_\op)\Bigr)u(\vec p , s)  \widetilde{\psi}(\vec p)\right\}.   \nonumber
\end{eqnarray}
If we assume that $\vec p$ and $\vec p\,'$ are both small on the scale over which $\overline{u}(\vec p\,',s') \vec{\mathcal{F}}^C u(\vec p , s) $ varies, this equation reduces to Eq.~\eqref{eq:moment_Xprod-2} since the factor
\begin{equation}
\sum_{\vec p, \vec p\,',\vec w} \widetilde{\psi'}(\vec p\,')^* e^{i\vec w \cdot(\vec p - \vec p\,')} \widetilde{\psi}(\vec p) = \langle \psi'|\psi\rangle,
\end{equation}
which can now be recognized on the right-hand side, cancels that on the left.

\section{Conventions}
\label{sec:conventions}

We adopt the following gamma matrix convention:
\begin{eqnarray}
  \sigma_x  =  \left(\begin{array}{cc}
    0 & 1\\
    1 & 0
  \end{array}\right) \quad
  \sigma_y  =  \left(\begin{array}{cc}
    0 & - i\\
    i & 0
  \end{array}\right) \quad
  \sigma_z  =  \left(\begin{array}{cc}
    1 & 0\\
    0 & - 1
  \end{array}\right).
\end{eqnarray}
\begin{eqnarray}
  \gamma_0  &=&  \left(\begin{array}{cc}
    0 & 1\\
    1 & 0
  \end{array}\right),\quad
  \gamma_1 =  - i \left(\begin{array}{cc}
    0 & \sigma_x\\
    - \sigma_x & 0
  \end{array}\right),\quad
  \gamma_2  =  - i \left(\begin{array}{cc}
    0 & \sigma_y\\
    - \sigma_y & 0
  \end{array}\right), \nonumber\\
  \gamma_3 &=&  - i \left(\begin{array}{cc}
    0 & \sigma_z \\
    - \sigma_z & 0
  \end{array}\right), \quad
  \gamma_5 = \left(\begin{array}{cc}
    1 & 0\\
    0 & - 1
  \end{array}\right) = \gamma_0 \gamma_1 \gamma_2 \gamma_3.
\end{eqnarray}
The continuum fermion propagator is
\begin{eqnarray}
  S(x,y) &=& \int \frac{d^4 p}{(2\pi)^4} \frac{1}{i\slashed p+m} e^{i p \cdot (x-y)}.
\end{eqnarray}
The two Dirac positive-energy, zero-momentum eigenstates are
\begin{eqnarray}
  u(\vec p = \vec 0, s) &=& \frac{1}{\sqrt{2}}\left(\begin{array}{c}
    \chi_s \\
    \chi_s
  \end{array}\right),   \quad
  \overline{u}(\vec p = \vec 0, s) = \frac{1}{\sqrt{2}}\left(\begin{array}{cc}
    \chi_s^\dagger & \chi_s^\dagger
  \end{array}\right),
\end{eqnarray}
where
\begin{eqnarray}
  \chi_0 = \left(\begin{array}{c}
    1 \\
    0
  \end{array}\right), \quad
  \chi_1 = \left(\begin{array}{c}
    0 \\
    1
  \end{array}\right).
\end{eqnarray}

\bibliography{ref}

\end{document}